\documentclass{article}
\usepackage{PRIMEarxiv}

\usepackage[utf8]{inputenc} % allow utf-8 input
\usepackage{hyperref}       % hyperlinks
\usepackage{url}            % simple URL typesetting
\usepackage{booktabs}       % professional-quality tables
\usepackage{amsfonts}       % blackboard math symbols
\usepackage{nicefrac}       % compact symbols for 1/2, etc.
\usepackage{microtype}      % microtypography
\usepackage{lipsum}
\usepackage{fancyhdr}       % header
\usepackage{graphicx}       % graphics
\graphicspath{{media/}}     % organize your images and other figures under media/ folder
\usepackage{float} 

\usepackage{graphicx}
\usepackage[most]{tcolorbox}
\usepackage{caption}
\usepackage{subcaption}
\usepackage{longtable}

\usepackage{amsmath,amssymb,amsfonts}

\usepackage{textcomp}

\usepackage{verbatim}

\usepackage{xcolor}
\usepackage{url}
\usepackage{xr-hyper}
\usepackage{hyperref}
\usepackage{pdfpages} 
\usepackage{bm}
\usepackage{multirow}
\usepackage{multicol}
\usepackage{array}
\newcolumntype{L}{>{\arraybackslash}m{3cm}}

\usepackage{fullpage}
\usepackage{rotating}
\usepackage{stmaryrd}
\usepackage{proof}

\usepackage{tikz}
\usetikzlibrary{bayesnet}
\usetikzlibrary{decorations.pathmorphing} % noisy shapes
\usetikzlibrary{fit}% fitting shapes to coordinates
\usetikzlibrary{backgrounds}	

\usepackage{amsthm}

\usepackage[ruled]{algorithm2e}
\usepackage{algorithmic} %% Used in Chapter 1 

\usepackage[backend=biber, style=apa]{biblatex}
\title{Online learning techniques for prediction of temporal tabular datasets with regime changes}
\author{
  Thomas Wong \\
  Department of Mathematics \\
  Imperial College London \\
  South Kensington Campus \\
  London, \,  SW7 2AZ\\
  United Kingdom \\
  \texttt{mw4315@ic.ac.uk} \\
   \And
  Mauricio Barahona \\
  Department of Mathematics \\
  Imperial College London \\
  South Kensington Campus \\
 London, \,  SW7 2AZ\\
  United Kingdom \\
  \texttt{m.barahona@imperial.ac.uk} \\
}

\begin{document}

\maketitle

% \TW{Publication: Royal Society Open Science, 
% Machine Learning with Applications} 

% REQUIRED
\begin{abstract}
The application of deep learning to non-stationary temporal datasets can lead to overfitted models that underperform under regime changes. In this work, we propose a modular machine learning pipeline for ranking predictions on temporal panel datasets which is robust under regime changes. The modularity of the pipeline allows the use of different models, including Gradient Boosting Decision Trees (GBDTs) and Neural Networks, with and without feature engineering. We evaluate our framework on financial data for stock portfolio prediction, and find that GBDT models with dropout display high performance, robustness and generalisability with reduced complexity and computational cost. We then demonstrate how online learning techniques, which require no retraining of models,  can be used post-prediction to enhance the results. First, we show that dynamic feature projection
%, an efficient procedure that can be applied to any machine learning model, 
improves robustness by reducing drawdown in regime changes. Second, we demonstrate that dynamical model ensembling based on selection of models with good recent performance leads to improved Sharpe and Calmar ratios of out-of-sample predictions. We also evaluate the robustness of our pipeline across different data splits and random seeds with good reproducibility. 
\end{abstract}

% REQUIRED

% REQUIRED
% \begin{AMS}
    
% \end{AMS}

% \MB{Cite reviews and papers about ML methods in finance. You should be citing the key papers in this field even if it is just to say what has been done} 
% \MB{Also, you need to highlight in the introduction those topics that you have concentrated most with your tweaks in the algorithms --- feature projection $\to$ dynamic feature projection $\to$ the fact that GBM methods are better than NN methods $\to$ dynamical model selection, does it add anything? The intro has to highlight the problem, the approach and the main findings at a high level} 
% \MB{You are approaching here issues related to robustness for time-series tabular data --- explain what the important issues are specifically for such systems.} 
% \MB{Also there is the issue of retraining models that you also explore, and how to avoid it... This should be raised already.} 

% \MBB{In this particular field, the key issues are...  you need to say what these are:  temporal data, what else?} \\  
% \MBB{Then you also need to say what the main tasks are in this area and what methods have been used for those...} \\ 
% \MBB{Once you do that, you can then explain the challenges of robustness, online training, etc.   --- That way you focus the problem on what you are solving here.} 

\section{Introduction}
\label{sec:introduction}

As investors explore new ways to generate profit, machine learning (ML) models are increasingly used as part of trading strategies, e.g., to predict the future return of stocks or stock portfolios. 
In particular, deep learning models for time-series data, such as Recurrent Neural Networks (RNNs) and Convolutional Neural Networks (CNNs), have been applied to the prediction of future stock returns~\autocite{9469554,8126078,8355458}. However, a major challenge for such methods is the highly stochastic, non-stationary and non-ergodic nature of financial data, which violates the assumptions of many algorithms. 
%%%%  
Furthermore, deep learning models are over-parameterised, with numbers of parameters orders of magnitude larger than typical sizes of time series data. 
Therefore, deep models can be easily overfitted to specific patterns in historical market data not present in future market data, and the overfitting worsens with more complex neural network architectures, such as Long Short Term Memory (LSTM) or Transformer networks. In addition, the continuous influx of data, coupled with possible regime changes, requires costly updating and retraining of such models. As a consequence, such methods can lack reproducibility and robustness for the prediction of future market data. 

As pointed out in recent reviews \autocite{Hullman22, Kapoor22, Pham20, Gundersen23,gundersen2023sources,}, replication of ML studies is often difficult due to several issues, including data leakage \autocite{Kapoor22}, program bugs \autocite{Landos21,}, data and code usability \autocite{Samuel20,}, model representation and evaluation \autocite{Hullman22,}.
These problems 
%have varying impact on the reliability and robustness of ML models,cand 
are currently hindering the usage of ML in high-risk decision processes, such as healthcare and finance.
%%%%%  Data leakage
For financial trading applications in particular, these issues  can have critical effects on the validity of results. Data leakage, in the form of look-ahead bias or overlap of training/test sets~\autocite{BOLLEN20111}, can inflate in-sample performance with poor performance when deployed live. 
%%%% Robustness of methods
Furthermore, black-box ML models, such as neural networks, can lack robustness as they are highly sensitive to small changes in parameters, random seeds and data, thus resulting in highly variable predictions.
%%%  Update of models to incoming data
The non-stationarity of data and the presence of regime changes also mean that ML models need to be re-trained with the latest financial data, a task that is not only computationally costly but also introduces further uncertainty. Yet most studies do not consider model performance when trained on different segments of historical market data \autocite{9469554,8126078,8355458,9591781,8442645}. 
Although reinforcement learning (RL) in online learning settings allows ML models to adapt to changing environments, deep reinforcement learning models are complex and require large computational resources~\autocite{Moritz17}. Indeed, applying RL to stock trading is challenging since the complexity of the action space increases exponentially with the number of stocks in the portfolio.

The above issues suggest the need to develop robust ML pipelines for trading applications possibly based on simpler models that can still operate on non-stationary, highly stochastic data under regime changes. 
Here we consider such a pipeline for \textit{temporal tabular data}, which allows the use of traditional ML models, such as Gradient Boosting Decision Trees (GBDTs) and other ensemble methods, in trading different assets classes, such as individual stocks and exchange traded funds (ETFs) at different frequencies \autocite{Prado2020,ZHOU2019105747,Jingwen23}. 
%
%This approach could also allow the integration of additional sources of data, such as sentiment analysis of news articles to improve the prediction accuracy of the direction of stock returns \autocite{7550882}, although we have not pursued this line of work here.
%
In particular, we find that Gradient Boosting models, which are known to be robust to data perturbations, outperform neural network models in our tasks.
We also show that improved robustness of ML models and adaptation to regime changes can be attained 
%without the use of deep reinforcement learning 
by employing: (i) dynamic feature projection, a simple approach that reduces the linear correlation to a subset of features evolving in time, and (ii) dynamic model selection, a simple optimisation procedure that selects optimal models from an ensemble based on recent performance. These approaches robustly improve trading performances by reducing volatility and drawdown during adversarial market regimes.

\begin{figure}
    \centering
    \resizebox{\textwidth}{!}{
    \includegraphics{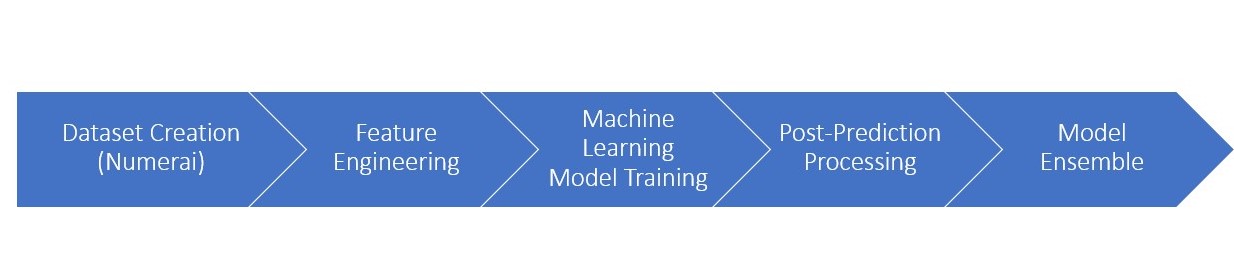}
    }
    \caption{\textbf{Schematic of the Machine Learning pipeline.} Starting with the Numerai data set, we consider feature engineering methods to augment the dataset and train an ML model (several are evaluated, including neural networks, but we settle for gradient boosting trees) to obtain the raw predictions. These then go through post-prediction processing (e.g., dynamic feature projection) to provide normalised predictions, which are then combined through model ensembling and dynamic model selection methods to output the predictions that are submitted to the Numerai tournament. 
    }
    \label{fig:robustML_pipeline}
\end{figure}

%%%%%%  Non transparent models
To exemplify the above issues, we consider a benchmark financial data platform that is continuously updated and curated under the Numerai tournament of stock portfolio prediction~\autocite{numerai}.
% Why use Numerai as an example 
Numerai is a hedge fund that organises a data science competition (as of May 2023) and provides free, open-source, high quality standardised financial data to all participants. 
As discussed below in more detail, the data set is given in the form of pre-processed temporal tabular data and the task is the prediction of the relative performances of stocks within an evolving trading universe without access to the identity of individual stocks.
Unlike other financial research papers that use proprietary data sets which can be difficult to validate~\autocite{8442645,9591781}, 
this open financial data competition allows researchers to replicate findings transparently and allows us to focus on establishing ML end-to-end pipelines which achieve consistent profits trading a market-neutral portfolio under changing market regimes. Our pipeline, shown in Figure~\ref{fig:robustML_pipeline}, is built upon simple, yet robust methodologies that avoid some of the problems of overfitting and high computational cost inherent to deep methods.
To enhance the robustness of the pipeline, each step is implemented independently avoiding data leakage, so that the pre-processing and actual model do not share data. Key ingredients of our pipeline are the post-prediction processing, based on online learning techniques, which allow easy adaptation of models to regime changes without expensive retraining.

% Structure of paper 
The paper is organised as follows. Section~\ref{section:numerai-data} introduces the Numerai datasets used in this paper and the prediction tasks. Section ~\ref{section:methods} describes and discusses the different computational methods, including online cross-validation, feature engineering and the different ML models considered and evaluated for the pipeline. Section~\ref{section:numerai-benchmark} presents the results from our ML pipeline, including the impact of different design choices on the robustness of trading performance. Performances of ML models under different market regimes are discussed in Section~\ref{section:numerai-regime}. In Section~\ref{section:online-learning}, we introduce adaptations based on online learning approaches, which help deal with regime changes, noting that these adaptations are generic and not limited to specific ML models. Lastly, we provide a summary and discussion, pointing open directions and alternatives together with a study of the robustness of our ML pipeline in Section~\ref{section:robustness}.

%\MB{If these are the steps of the pipeline, then these should appear somewhere in the text to organise the explanation below. And then you need to show where your contributions lie, i.e., which parts of the pipeline you are defining differently and what choices you make in each of the steps. } 

\section{Numerai dataset and prediction task} 
\label{section:numerai-data}

% Definition of temporal tabular data 
%\TW{Add a picture in case reviewers do not understand?} 

Financial data are often available in the form of time series. These time series can be treated directly using classic methods such as ARIMA models \autocite{percival_walden_2020} and more recently through deep learning methods such as Temporal Fusion Transformers \autocite{Bryan19}. 
However, such methods are easily overfitted and need expensive retraining for financial data, which are inherently affected by regime changes and high stochasticity. Alternatively, one can use feature engineering methods to transform these time series into \emph{tabular form} through a process sometimes called `de-trending' in the financial industry, where the characteristics of a financial asset at a particular time point, including features from its history, are represented by a data vector. In this representation, the time dimension is not considered explicitly, as the state of the system is captured through transformed features recomputed at each time point and the continuity of the temporal dimension is not used. 
For example, we can summarise the time series of the return of a stock with the mean and standard deviation over different look-back periods. Grouping these data rows for different financial assets into a table at a given time point we obtain a \emph{tabular dataset}. If the features are informative, this representation can be used for prediction tasks at each time point, and allow us to employ robust and widely tested ML algorithms that are applicable to tabular data. The Numerai competition is based on a curated tabular data set with high-quality features made available to the research community. Other recent examples of open source financial datasets can be found in recent work by Zimmermann et al. \autocite{chen2021open,}.

\paragraph{Description of the dataset:} The Numerai dataset is a temporal tabular dataset. A temporal tabular dataset is a collection of matrices $\{ X_i \}_{1 \leq i \leq T}$ collected over time eras 1 to $T$. Each matrix $X_i$ represents data available at era $i$ with shape $N_i \times M$, where $N_i$ is the number of data samples in era $i$ and $M$ is the number of features describing the samples. The definition of the features is fixed throughout the eras, in the sense that the same computational formula is used to compute the features in each week, whereas the number of data samples $N_i$ does not have to be constant across time. In the Numerai dataset considered here, the matrices $X_i$ contain $M$ stock market features (computed by Numerai) for $N_i$ stocks, which are updated weekly (i.e., in our case the eras are weeks). 

It is important to remark that the dataset is \textit{obfuscated}, 
%so that the proprietary data generation process from financial datasets is not known to the participants. It is 
i.e., it is not possible for participants to know the identity of stocks or even which stocks are present each week. Each data row is indexed by a hash index, known only to Numerai, that maps the data rows to the stocks. As a result, it is not possible for participants to concatenate different data rows to create a continuous history of a stock. The matrix $X_i$ thus provides a snapshot of the market at week $i$ presented as an unknown set of stocks described by a common set of features, such that the features are computed consistently across all stocks in the week and also computed consistently across different weeks.      

The Numerai dataset starts on 2003-01-03 (Era 1). The tabular set has 1181 features, which are already normalised into 5 equal-sized integer bins, from 0 to 4. (Note that an initial release of the dataset had 1191 features but Numerai removed 10 ill-defined features in a subsequent refinement.) There are 28 target labels which are derived from stock returns using 14 proprietary normalisation methods (nomi, jerome, janet, ben, alan, paul, george, william, arthur, thomas, ralph, tyler, victor, waldo) each over 2 forward-looking periods (20 trading days, 60 trading days). The main target label to evaluate performance is target-nomi-v4-20, i.e., forward 20 trading days return obtained by the nomi normalisation method.
%Other targets are named similarly. The target labels are all 
This target label is scaled between 0 and 1, where a smaller value represents a lower forward return, and grouped into 5 bins (0, 0.25, 0.5, 0.75, 1.0)
%For each normalisation method, the number of bins could be different, 5 to 7 bins are created for each target with the bin sizes 
following a Gaussian-like distribution. 
%so that most stocks are within the central bin of 0.5 while only a small amount of stocks are in the tail bins of 0 and 1. 
For our analysis, we transform the features and labels so that both become zero-mean. For features, we subtract 2 from the integer bins so that the transformed bins are -2,-1,0,1,2. For the target labels, we subtract 0.5 so that the transformed targets are in the range -0.5 to 0.5). 

\paragraph{Prediction task:} The task in the Numerai tournament is to predict the \emph{stock rankings} each week, ordered from lowest to highest expected return. The scoring is based on Spearman's rank correlation %(defined as \textbf{Corr}) 
of the predicted rankings with the main target label (target-nomi-v4-20). Hence there is a single overall score each week regardless of the number of stocks to predict each week. Participants are not scored on the accuracy of the ranking of each stock individually, only on the overall score. 
Numerai uses the predicted rankings to construct a market-neutral portfolio which is traded every week (as of Dec 2022), i.e., the hedge fund buys and short-sells the same dollar amount of stocks. Therefore the relative return of stocks is more relevant than the absolute return, hence the prediction task is a ranking problem instead of a forecasting problem. 

\section{Methods}

\label{section:methods}

\subsection{Robustness in Machine Learning pipelines} 

In this paper, we aim to design an ML pipeline for temporal tabular datasets focusing on its robustness. Table~\ref{table:data-leakage} details issues related to robustness and reproducibility, as listed in a recent review \autocite{Kapoor22,}, and how they are addressed in our paper. By preventing look-ahead bias and other data leakage issues, our pipeline can be robustly applied to live trading setups. 

In addition to avoiding data leakage, the following design choices are used to improve the robustness and reliability of the results. Firstly, the impact of random seeds is reduced by reporting results from average predictions over 10 different random seeds for each machine learning method. Secondly, the metrics used for model evaluation are the same as in the Numerai tournament to avoid researcher bias in discounting unfavourable results. Finally, cross-validation is independent of random seeds and any other human selection, thus reducing the chance of overfitting models to a particular data split.

\begin{table}[htb!]
%\begin{tabular}{|l|l|}
\begin{tabular}{|p{.3\linewidth}|p{0.62\linewidth}|}
\hline
\textit{Issues affecting robustness of ML algorithms} &     \textit{How the issue is addressed here}  \\ 
\hline \hline
`No test set'  & 
A robust cross-validation scheme is used. \\ 
\hline
`Pre-processing on training and test set'  & 
Numerai features are already standardised; hence minimal pre-processing. \\ 
\hline
`Feature selection on training and test set' &
Feature Engineering is applied to each data row independently\\ 
\hline
`Duplicates in datasets' &
A unique id for each data row reduces the chance of duplicates in dataset \\ 
\hline
`Model uses features that are not legitimate' &
Only data provided by Numerai is used to train ML models---no extra features from other resources, and no cherry-picking of features. \\ 
\hline
`Temporal leakage' &
 We use \textit{Grouped Time-Series Cross-Validation} with no overlap between training/validation/test (Fig.~\ref{fig:ts-crossvalidation}). \\
 & Feature Engineering is applied to each data row independently, i.e., no data leakage between eras. \\ 
 \hline
`Non-independence between training and test' &
Training and test samples are market data at different periods without overlap. \\ 
\hline
`Sampling bias in test distribution' &
The stocks trading each week are decided by Numerai based on operational and risk considerations.  \\ 
\hline
\end{tabular}
\caption{\textbf{Data analysis design.}  Some common issues regarding data leakage in machine learning research~\autocite{Hullman22, Kapoor22, Pham20,} and how these issues are dealt with in this study.}
\label{table:data-leakage}
\end{table}

% \subsubsection{Grouped Time-Series Cross-Validation} 

% \label{section:tscv}

%For temporal tabular datasets, we use a new form of cross-validation, grouped time-series cross-validation. 
For temporal datasets, standard cross-validation schemes 
%such as grouped cross-validation 
cannot be used directly, as a random split of eras could lead to the training set including data that appears later in time than the validation and test sets, hence introducing look-ahead bias.  To avoid this problem, we use \emph{grouped time-series cross-validation}, which splits eras according to their chronological order (Fig.~\ref{fig:ts-crossvalidation}).  Since for financial applications the target labels often involve future asset returns and are reported with a lag, we add a gap between the training and validation sets and similarly between the validation and test sets.

%% Diagram of showing grouped time-series cross-validation. 

\begin{figure}[bth!]
\centering
\resizebox{.8\linewidth}{!}{
\tikzset{every picture/.style={line width=0.75pt}} %set default line width to 0.75pt        
\begin{tikzpicture}[x=0.75pt,y=0.75pt,yscale=-1,xscale=1]

%Shape: Rectangle [id:dp6647729752887366] 
\draw   (11,69) -- (225,69) -- (225,109) -- (11,109) -- cycle ;
%Shape: Rectangle [id:dp738309048429999] 
\draw   (300,124) -- (439,124) -- (439,164) -- (300,164) -- cycle ;
%Shape: Rectangle [id:dp6587860593804173] 
\draw   (524,176) -- (654,176) -- (654,216) -- (524,216) -- cycle ;
%Straight Lines [id:da8693874586257924] 
\draw    (475,171) -- (516,171) ;
\draw [shift={(518,171)}, rotate = 180] [color={rgb, 255:red, 0; green, 0; blue, 0 }  ][line width=0.75]    (10.93,-3.29) .. controls (6.95,-1.4) and (3.31,-0.3) .. (0,0) .. controls (3.31,0.3) and (6.95,1.4) .. (10.93,3.29)   ;
%Straight Lines [id:da7882322969133861] 
\draw    (475,171) -- (448,171) ;
\draw [shift={(446,171)}, rotate = 360] [color={rgb, 255:red, 0; green, 0; blue, 0 }  ][line width=0.75]    (10.93,-3.29) .. controls (6.95,-1.4) and (3.31,-0.3) .. (0,0) .. controls (3.31,0.3) and (6.95,1.4) .. (10.93,3.29)   ;

%Straight Lines [id:da25952138105822553] 
\draw    (256,117) -- (297,117) ;
\draw [shift={(299,117)}, rotate = 180] [color={rgb, 255:red, 0; green, 0; blue, 0 }  ][line width=0.75]    (10.93,-3.29) .. controls (6.95,-1.4) and (3.31,-0.3) .. (0,0) .. controls (3.31,0.3) and (6.95,1.4) .. (10.93,3.29)   ;
%Straight Lines [id:da8569047524286293] 
\draw    (256,117) -- (229,117) ;
\draw [shift={(227,117)}, rotate = 360] [color={rgb, 255:red, 0; green, 0; blue, 0 }  ][line width=0.75]    (10.93,-3.29) .. controls (6.95,-1.4) and (3.31,-0.3) .. (0,0) .. controls (3.31,0.3) and (6.95,1.4) .. (10.93,3.29)   ;

%Straight Lines [id:da9661891556407889] 
\draw    (10,230) -- (670,230) ;
\draw [shift={(668,230)}, rotate = 180] [color={rgb, 255:red, 0; green, 0; blue, 0 }  ][line width=0.75]    (10.93,-3.29) .. controls (6.95,-1.4) and (3.31,-0.3) .. (0,0) .. controls (3.31,0.3) and (6.95,1.4) .. (10.93,3.29)   ;

% Text Node
\draw (65,79.4) node [anchor=north west][inner sep=0.75pt]    {$Training\ Data$};
% Text Node
\draw (315,138.4) node [anchor=north west][inner sep=0.75pt]    {$Validation\ Data$};
% Text Node
\draw (551,187.4) node [anchor=north west][inner sep=0.75pt]    {$Test\ Data$};
% Text Node
\draw (290,238) node [anchor=north west][inner sep=0.75pt]   [align=left] {Time};
% Text Node
\draw (250,98) node [anchor=north west][inner sep=0.75pt]   [align=left] {gap};
% Text Node
\draw (470,150) node [anchor=north west][inner sep=0.75pt]   [align=left] {gap};
\end{tikzpicture}
}
\caption{Illustration of data split using grouped time-series cross-validation}
\label{fig:ts-crossvalidation}
\end{figure}

\subsection{Feature Engineering (FE) methods} 

%Feature engineering is an important step to enhance the power of tabular methods for the analysis of time series data. 
We evaluate different feature engineering methods that can be applied to temporal tabular data sets with numerical features, as applicable to the Numerai data set. 
% Polynomial Transformation of features
%\paragraph{Polynomial Transformations} 
Firstly, new features can be created by applying polynomial transformations, such as multiplication and addition, to the original features. Here we create new features by multiplying two features, which can be thought of as capturing interactions between feature pairs. To alleviate the computational cost, when the number of features is large, we create new features from a random subset of feature pairs. Note that the computation of these features can be done in parallel for data from each era. 
% drop-out 
%\paragraph{Dropout} 
Secondly, a simple way of data augmentation is to add randomness to the feature matrix with different dropout methods, which are used extensively to reduce overfitting in neural network models \autocite{Arlind21}.
Here we apply dropout by multiplying the original data with a Boolean mask so that some numerical features are set to zero. The dropout is characterised by its sparsity level (how many features are set to zero) and its sparsity structure (how to choose the features set to zero). Since our tabular dataset has no local spatial structure, we use a random Boolean matrix with uniform probability. This encourages the ML methods to learn multiple feature relationships and reduces reliance on a small set of important features.

%% Feature Engineering that we use 
For our dataset, we first augment the feature matrix by creating additional features by multiplying feature pairs, followed by dropout with a random Boolean mask on the augmented feature matrix. A grid search is used to find optimal hyperparameters of the feature engineering methods, i.e., the number of feature products and the sparsity of the dropout.

\subsection{Machine Learning algorithms for tabular datasets}
%% Current issue: What ML models to use for tabular dataset 
%\paragraph{Choosing Machine Learning methods} 
Numerous machine learning models have been proposed for tabular datasets, and different benchmarking studies have shown conflicting views on their performance~\autocite{Shwartz21, Arlind21,}. The biggest disagreement in the literature is whether gradient-boosting decision trees or neural networks are superior in regression and classification tasks of tabular datasets. Whereas one paper claims gradient boosting models (XGBoost) outperformed deep learning models in 8 out of 11 datasets and none of the deep learning models consistently outperform others \autocite{Shwartz21}, another paper suggests that well-tuned multi-layer perceptron (MLP) models with regularisation can outperform different gradient boosting models such as XGBoost and CatBoost \autocite{Arlind21,}. Interestingly, both these studies share the view that neural networks with complicated designs, such as attention layers and other transformer layers, tend to generalise poorly, with a strong drop in performance when applied to data sets beyond their original study. 
%%% How we design benchmarking study 
%
Importantly, the Numerai data set is different from the data sets in the above benchmarking studies in that it is incremental instead of fixed. Hence the data distribution varies across time periods due to market regime effects, and we do not have a homogeneous distribution across cross-validation splits. With such a different problem setup, it is thus not possible to use the above benchmarking studies to guide our choice of ML method.  

In this study, we benchmark a range of machine learning models, including different variants of gradient-boosting decision tree models and neural network models. The choice of ML models is based on the popularity of usage in data science competitions and code quality, as one of our aims is replicability of results. We train all machine learning models with a single GPU, the standard setup for most participants in data science competitions. Some brief details of the ML models used are provided in the following. 

% What is Gradient Boosting 
%\paragraph{Gradient Boosting} 
\paragraph{Gradient Boosting Decision Trees (GBDTs):}

Boosting can be seen as an extension of generalised additive models (GAM) where the additive components of smooth parametric functions can be replaced by any weak learners such as decision trees \autocite{Kotsiantis2011DecisionTA}. Historically, various boosting algorithms have been proposed for different loss functions. For example, AdaBoost \autocite{FREUND1997119} was proposed for binary classification problems with exponential loss, whereas Gradient Boosting was first proposed by Friedman in 2001 \autocite{FriedmanJeromeH.2001GfaA} for any smooth loss functions. Algorithm \ref{alg:gradient-boosting} in the SI outlines the iterative update equations of gradient boosting. 

Of the various Python implementations of GBDTs, we use LightGBM \autocite{LightGBM} in this paper. CatBoost \autocite{CatBoost} is not used here as the Numerai dataset has no categorical features.

Algorithm \ref{alg:gradient-boosting-tree} in the SI shows how LightGBM implements gradient boosting with decision trees being the weak learners, including several computational and numerical improvements relative to XGBoost and other implementations. In addition to traditional gradient boosting decision trees (\textit{LightGBM-gbdt}), we consider two other implementations of GBDT models:
\begin{itemize}
%% Modifications by dart and goss 
\item \textit{Dropouts meet Multiple Additive Regression Trees (LightGBM-dart)} ignores a portion of trees when computing the gradient for subsequent trees \autocite{pmlr-v38-korlakaivinayak15}, thus avoiding over-specialisation where the later learned trees can only affect a few data instances. This reduces the sensitivity of models towards decisions made by the first few trees. 
\item \textit{Gradient-based One-Side Sampling (LightGBM-goss)} reduces the number of data instances used to build each tree: it keeps data instances with large absolute gradients and randomly samples a subset of data with small absolute gradients. The approximation error of the gradient using LightGBM-goss converges to the standard method when the number of data is large, and it outperforms other data sampling (e.g., uniform sampling) in most cases. 
\end{itemize}

%% Implementations of GBDT in LightGBM
For all LightGBM models, we use mean squared error (L2 loss) as the loss function for the regression problems. The number of trees and learning rate is optimised by hyperparameter searches. To prevent overfitting, the maximum depth and number of leaves in each tree and the minimal number of data samples in the leaves are tuned for each model. L1 and L2 regularisation are also applied. Data and feature sub-sampling are used to reduce similarities between trees: before building each tree, a random part of data is selected without re-sampling and a random subset of features is chosen to build the tree. For LightGBM-dart models, both the probability to apply dropout during the tree-building process and the portion of trees to be dropped out are tuned. Early stopping is applied using the validation dataset for LightGBM-gbdt models to further reduce overfitting.

\paragraph{Neural Networks (NNs):}  

The most basic architecture of neural networks, multi-layer perceptron (MLP), has failed to outperform gradient boosting models in benchmark studies of tabular datasets \autocite{Shwartz21}. 
Recently, more complex network architectures have been proposed for tabular data sets, as surveyed in ~\autocite{Borisov21,} 
These new architectures can be classified into two major groups: 
 \begin{itemize}
     \item \textit{Hybrid models} that combine neural networks with other traditional ML methods, e.g., decision trees. 
     Neural Oblivious Decision Ensembles (NODE) \autocite{NODE} is a generalisation of gradient boosting models into differentiable deep decision trees allowing end-to-end training with gradient descent optimisers such as PyTorch \autocite{PyTorch}. DeepGBM \autocite{DeepGBM} combines two neural networks: CatNN to handle sparse categorical features, and GBDT2NN to distil tree structures from a pre-trained GBDT model to handle numerical features. 
     A major limitation of these models is the large memory consumption, 
     which makes them run out of memory on the NVIDIA 3080ti GPU. Therefore we do not use them in our benchmarking here.
     \item \textit{Transformer-based models} that use deep attention mechanisms to model complex feature relationships. TabNet \autocite{Arik_Pfister_2021} uses sequential attention to perform instance-wise feature selection at each decision step, enabling interpretability and better learning. AutoInt \autocite{AutoInt_Song_2019} maps and models feature interactions in a low-dimensional space with a multi-head, self-attentive neural network with residual connections. AutoInt runs out of memory on a single GPU and is thus not used in our benchmarking. Tabnet also has similar memory issues, hence we down-sampled the data by keeping every fifth week of data (i.e., 20\% of the original data) for the training/validation periods, so that Tabnet could be trained on the single GPU used in this study. Our aim is to compare performance under modest computational resources attainable by a wide class of users. 
 \end{itemize}
 
In summary, our benchmark analysis includes two NN models: MLP and TabNet implemented in PyTorch. We use mean squared error (L2 loss) for the regression problems.

\section{Evaluation of Machine Learning methods for the Numerai temporal tabular data set} 
\label{section:numerai-benchmark}

In this section, we study different ML methods applied to the Numerai temporal tabular data set for the prediction of stock rankings aimed at market-neutral stock portfolios.

\paragraph{Data Split} 
We use the latest version (v4) of the Numerai dataset.
The training period is fixed between 2003-01-03 (Era 1) to 2012-07-27 (Era 500); the validation period is fixed between 2012-12-21 (Era 521) and 2014-11-14 (Era 620); and the test period starts on 2015-05-15 (Era 646) and ends on 2022-09-23 (Era 1030). We apply a 1-year gap between training and validation periods to reduce the effect of recency bias so that the performance of the validation period will better reflect future performance. The gap between the validation period and test period is set to 26 weeks to allow for sufficient time to deploy trained machine learning models.

%% Evaluation 
\paragraph{Evaluation of performance} For each configuration of each ML method, we average over the predictions of different targets before scoring. The predictions are scored in each era by calculating the correlation (\textbf{Corr}) between the rank-normalised predictions and the actual (binned) stock ranking. The mean and standard deviation (volatility) of \textbf{Corr} are reported for both the validation and test periods. To measure the downside risk of the model, we also compute the \textit{Maximum Drawdown}, defined as the largest drop suffered by an investor starting at any time during the validation/test period.
As summary measures, we compute two standard ratios: 
(i) the Sharpe ratio, defined as the ratio of the mean and standard deviation of \textbf{Corr}; and (ii) the Calmar ratio, defined as the ratio of mean \text{Corr} over Maximum Drawdown. Good performance is characterised by large values of both of these ratios.

\paragraph{Model Training} 
We use Optuna~\autocite{Takuya19} to perform the hyperparameter search (see Section~\ref{section:optuna} in Supplementary Information) and select the hyperparameters with the highest Sharpe ratio for the main target (target-nomi-v4-20) in the validation period.  The optimised hyperparameters for each ML method are then fixed, and we train 10 models, starting the algorithms from 10 different random seeds. We report the average prediction of these 10 models for evaluation.

%% ML models 
\paragraph{Baseline Model} Momentum models are a common approach for the analysis of financial time series \autocite{stein2022modeling}. As a baseline, we consider a factor momentum model created by linear combinations of signed features, where the sign of each feature is determined by the sign of the 52-week moving average of correlations of that feature with the target.
This serves as a simple baseline \textit{linear} model, which is then compared to the ML models that can capture non-linearity in the data.

\paragraph{Comparative results of the different ML algorithms and Feature Engineering}
%% Tables 
Table~\ref{table:numerai-v4-benchmark} shows the performance of the different ML algorithms with and without feature engineering on: (a) the validation set, and (b) the test set. We focus on methods that achieve the highest mean \textbf{Corr}, and Sharpe and Calmar ratios.

\begin{table}[htb!]
(a) Performance over the validation period (2012-12-21 to 2014-11-14) 
\\
\begin{center}
\resizebox{.8\linewidth}{!}{
\begin{tabular}{|r|r|r|r||r|r|}
\hline
Method                & Mean   & Volatility & Max Draw & Sharpe & Calmar \\ \hline
Factor Momentum (baseline)   & 0.0229 & 0.0170     & 0.0691        & 1.3495 & 0.3314 \\ \hline
MLP with FE              & 0.0423 & 0.0208     & 0.0241        & 2.0338 & 1.7552 \\ 
MLP without FE           & 0.0443 & 0.0201     & 0.0065        & 2.2058 & \textbf{6.8154} \\ \hline
TabNet without FE        & 0.0362 & 0.0189     & 0.0199        & 1.9125 & 1.8191 \\ \hline
LightGBM-gbdt with FE    & 0.0483 & 0.0229     & 0.0307        & 2.1144 & 1.5733 \\ 
LightGBM-gbdt without FE & 0.0500 & 0.0224     & 0.0235        & \textbf{2.2335} & 2.1277 \\ \hline
LightGBM-dart with FE    & 0.0496 & 0.0223     & 0.0215        & \textbf{2.2274} & \textbf{2.3070} \\
LightGBM-dart without FE & 0.0475 & 0.0199     & 0.0079        & \textbf{2.3883} & \textbf{6.0127} \\ \hline
LightGBM-goss with FE    & 0.0288 & 0.0219     & 0.0687        & 1.3136 & 0.4192       \\ 
LightGBM-goss without FE & 0.0302 & 0.0234     & 0.0877        & 1.2877 & 0.3444 \\ \hline
\end{tabular}
}
% \label{table:numerai-v4-benchmark-validate}
%  \end{table}
\end{center}

%
% \begin{table}[htb!]
\bigskip
(b) Performance over the test period (2015-05-15 to 2022-09-23) 
\\
\begin{center}
\resizebox{.8\linewidth}{!}{
\begin{tabular}{|r|r|r|r||r|r|}
\hline
Method                     & Mean   & Volatility & Max Draw & Sharpe & Calmar \\ \hline
Factor Momentum (baseline) & 0.0080 & 0.0275     & 0.7877   & 0.2923 & 0.0102 \\ \hline
MLP with FE                & 0.0237 & 0.0330     & 0.2912   & 0.7189 & 0.0814 \\ 
MLP without FE             & 0.0258 & 0.0289     & 0.1668   & \textbf{0.8931} & \textbf{0.1547} \\ \hline
TabNet without FE          & 0.0161 & 0.0296     & 0.5811   & 0.5431 & 0.0277 \\ \hline
LightGBM-gbdt with FE      & 0.0253 & 0.0327     & 0.3064   & 0.7731 & 0.0826 \\ 
LightGBM-gbdt without FE   & 0.0262 & 0.0321     & 0.2378   & 0.8140 & 0.1102 \\ \hline
LightGBM-dart with FE      & 0.0265 & 0.0319     & 0.2151   & \textbf{0.8313} & \textbf{0.1232} \\ 
LightGBM-dart without FE   & 0.0278 & 0.0284     & 0.1622   & \textbf{0.9791} & \textbf{0.1714} \\ \hline
LightGBM-goss with FE      & 0.0169 & 0.0297     & 0.5539   & 0.5695 & 0.0305 \\ 
LightGBM-goss without FE   & 0.0156 & 0.0318     & 0.7528   & 0.4896 & 0.0207 \\ \hline
\end{tabular}
}
\end{center}

\caption{\textbf{Performance of different machine learning methods with and without feature engineering on the Numerai dataset for (a) validation period and (b) test period.} The three top methods according to Sharpe ratio and Maximum Drawdown over the validation period are shown in italics in (a). The top method according to the Sharpe ratio and Maximum Drawdown over the test period is shown in boldface in (b). For TabNet, the pipeline with feature engineering cannot be run due to memory constraints.}
\label{table:numerai-v4-benchmark}
\end{table}

%% Comparing Performance of different ML models
%\paragraph{Relative performance to baseline} 
Firstly, almost all ML models performed substantially better than the factor momentum model (baseline), in both validation and test periods. Whereas the factor momentum model relies on linear relationships, the capability of ML models to learn non-linear relationships, in addition to linear ones, adds to their robustness and improved performance under different, often volatile, market regimes.  

%\paragraph{Improvement from Feature Engineering}  
Secondly, we observe that Feature Engineering does not improve the performance of ML models. Although, in principle, Feature Engineering allows GBDT-based methods to model feature interactions more easily, our results suggest that these interactions are overfitted during the training process. For neural network-based models, feature engineering is not strictly necessary, as dropout is already embedded in network architectures.

%\paragraph{Selecting optimal ML models for test period} 
Thirdly, we note that all ML models scored better in the validation period than the test period. This is expected, as it is well known that the performance of trading models deteriorates over time due to overcrowding and regime changes (a phenomenon known as alpha decay). Models that are overfitted to recent training data will experience greater alpha decay than properly regularised models. 
%Therefore we select the ML method based on validation period performance through a two-step process: we first filter down to the top three models with the highest mean \text{Corr} in the validation period; and from those, we select the model with the highest Calmar ratio in the validation period. Selection by mean \text{Corr} ensures the model has good overall performance whereas the Calmar ratio considers the worst-case scenario, thus capturing the tail risks of the trading model. 
To select the ML method, we consider the top models according to the Sharpe and Calmar ratios over the validation period: a high Sharpe ratio ensures the model has good overall performance, whereas a high Calmar ratio ensures good performance against the worst-case scenario, thus capturing the tail risks of the trading model. 
Indeed, we find that \emph{LightGBM-dart without feature engineering} generalises well to the test period further into the future.

%\paragraph{Comparison of different ML models}  

% Comparing GBDT and NN 
%Finally, we find that Gradient Boosting models (LightGBM-gbdt/LightGBM-dart) have better generalisation ability than neural network-based models (MLP/TabNet), 
Finally, we note that LightGBM-gbdt has better generalisation to the test period than neural network-based models (TabNet), suggesting overfitting in these complex deep NN models. This indicates that although over-parameterised models can learn non-linear relationships in temporal tabular data sets, these relationships may be difficult to generalise under non-stationary data environments. 
Our results suggest that, despite their relative simplicity, gradient Boosting models capture non-linearity in a more robust and controlled manner, with early trees capturing linear relationships and non-linear relationships captured by the later trees, thus reducing the risk of catastrophic forgetting \autocite{doi:10.1073/pnas.1611835114}.

%\paragraph{Summary} 

In summary, we find that the best performing model in our dataset is LightGBM-dart without feature engineering. In the rest of the paper, we concentrate on this model to illustrate how the pipeline can be further improved with online learning to account for regime effects. 
However, to demonstrate the robustness of our pipeline and how it can be applied to improve the performance of any ML model, we will also report two other models: a similar GBDT model (LightGBM-gbdt without feature engineering) and a neural network model (MLP without feature engineering).

\section{Dealing with regime effects in the ML pipeline} 
\label{section:numerai-regime}

Financial data are heavily influenced by regime changes.  
Growth (`Bull') markets are characterised by low volatility and positive expected return, whereas high volatility and negative expected returns are characteristic of adverse (`bear') markets. Switches between regimes can be triggered by externalities, such as pandemics, economic recessions, etc.
From the perspective of the Numerai data set, such regime effects affect model performance. Volatility is detrimental to long-term performance due to the negative compounding of investment losses, a phenomenon known as `volatility tax'. Given that hedge funds are leveraged, we want to build consistent models with reasonably good performance under different market regimes, rather than models that have excellent performance in one market regime but fail in others.  

We now focus on regime effects when using ML models for financial tabular temporal data sets. Specifically, we consider the effect of feature projection, and reducing the dependence on the initial trees in gradient boosting models. 

%% NRVIX 
\begin{figure}[htb!]
    \centering
        \begin{subfigure}{0.45\textwidth}
        \includegraphics[width=\textwidth]{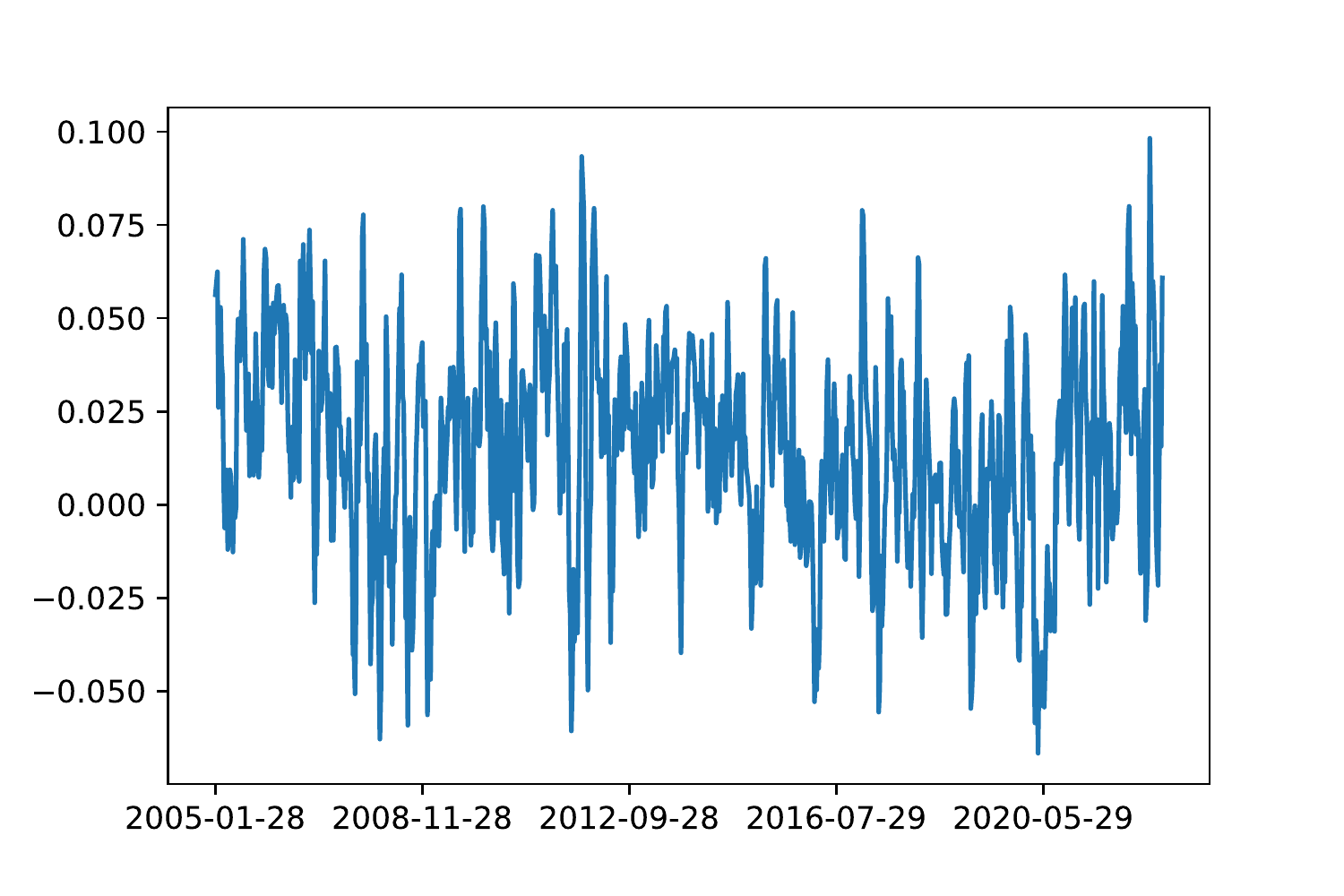}
        \caption{NMI}
        \label{fig:NRVIX-corr}
    \end{subfigure}
    \begin{subfigure}{0.45\textwidth}
        \includegraphics[width=\textwidth]{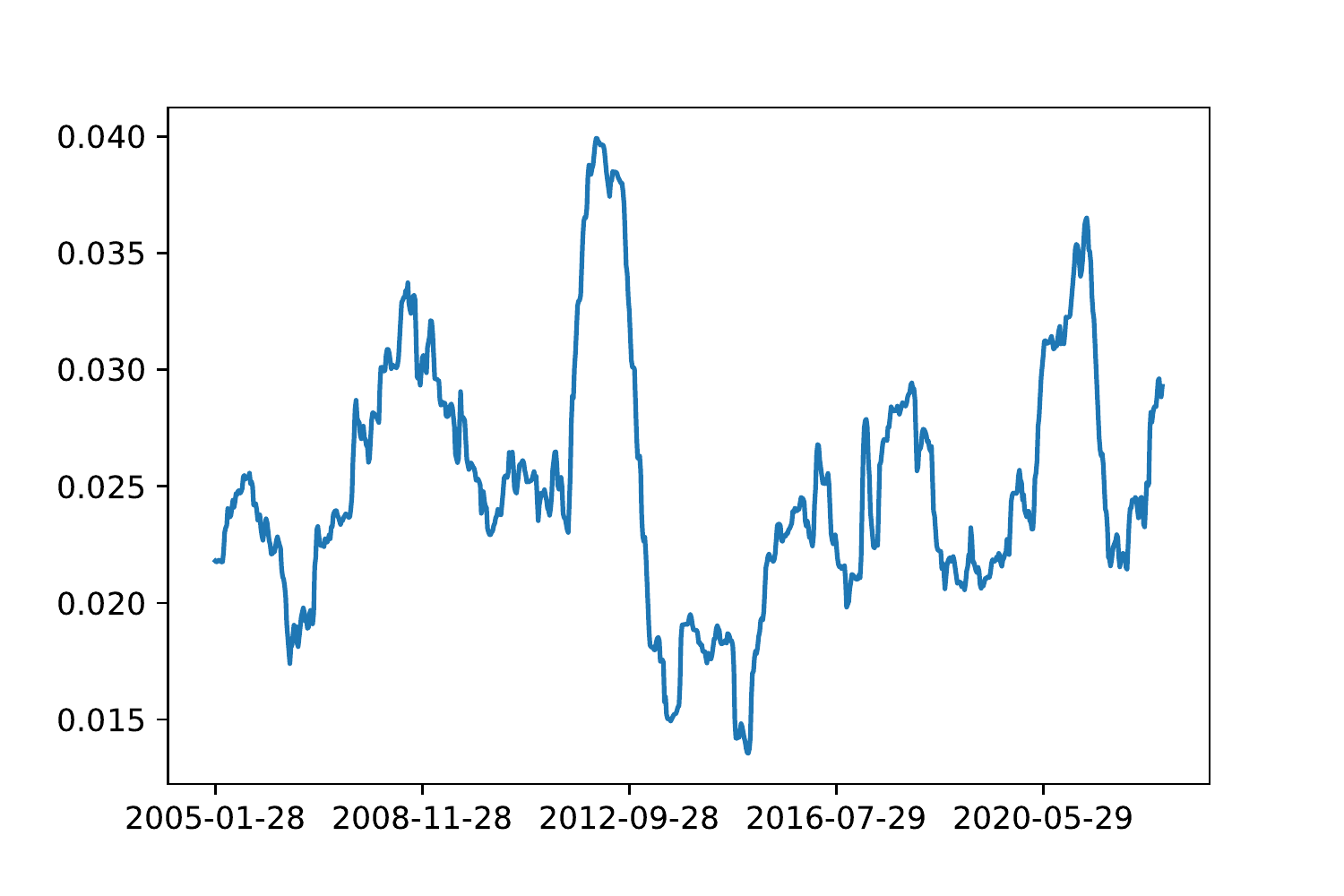}
        \caption{NRVIX}
        \label{fig:NRVIX-vix}
    \end{subfigure}
\caption{\textbf{High and low volatility regimes in the Numerai data.} (a) Numerai Market Index (NMI) for the period between 2005-01-28 (Era 109) and 2022-09-23 (Era 1016); (b) the computed Numerai Realised Volatility Index (NRVIX) used to identify the high and low volatility regimes. The high volatility regime refers to weeks where NRVIX is higher than 0.025 and the low volatility regime refers to weeks where NRVIX is lower than 0.025. }
\label{fig:NRVIX}
\end{figure}

\paragraph{Classification into high and low volatility regimes}
To classify the financial market into regimes, we consider an intrinsic measure derived directly from the Numerai dataset, as follows. We first compute the Numerai Market Index (NMI), i.e., the weekly performance of the baseline (linear) factor momentum portfolio,
%
% As shown below, the performance of baseline model, based on linear relationships in the dataset is a good indicator of performance of different ML models, which use both linear and non-linear relationships in the dataset. The performance of baseline model indicates the `difficulty' of a week for ML predictions, as linear relationships are easier to learn than non-linear ones. 
and we then calculate the Numerai Realised Volatility Index (NRVIX), defined as the standard deviation of NMI rolling over 52 weeks (Fig.~\ref{fig:NRVIX}). The eras are then classified into high and low volatility, based on a threshold of NRVIX=0.025, which is the mean over the first 7 years of data (2003-01-03 to 2010-02-26).  
%High NRVIX implies higher difficulty for ML predictions associated with high  volatility regimes. 
According to this intrinsic characterisation, low volatility regimes have stable linear relationships of features to stock returns, often associated with a good performance by ML models. On the other hand, high Volatility regimes correspond to unstable linear relationships of features to stock returns, which lead to poor model performance. 
Figure \ref{fig:NRVIX} shows that the high and low NRVIX regimes are well aligned with macroeconomic events: high volatility regimes include the financial crisis (2007-2009), the Euro crisis (2011-2012), and the Covid pandemic (2020), whereas low volatility regimes correspond to benign market conditions with no significant macro event risks, during which the factor momentum baseline portfolio had good returns.

% feature projection
\subsection{Feature Projection based on a fixed set of features}
\label{sec:standard_feature_projection}
%% Why perform feature projection 
Feature projection is the elimination of the effect of particular features in the model, thus reducing the risk of over-relying on individual features. Because the predictive ability of individual features is highly dependent on market regimes, it is undesirable to have ML models with heavy (linear)-dependence on certain features since this can lead to long periods of drawdown if there is a regime change.

%% Feature Neutral Correlation
We start by evaluating the feature projection suggested by the Numerai tournament. Numerai recommends that participants reduce model dependencies on 420 `risky features' (out of the 1181 features). 
This risky features are projected out by subtracting the linear correlation, as follows. Given a week of data with $n$ stocks, let $X \in \mathbb{R}^{n \times 420}$ be the matrix of risky features and $y \in \mathbb{R}^n$ the predicted rankings obtained from a model. For a given projection strength $\beta, 0 \leq \beta \leq 1$, the projected predicted ranking $\hat{y}$ is calculated as $ \hat{y} = y - \beta \, X X^\dagger y$, where $X^\dagger$ is the pseudo-inverse of $X$. The Feature Neutral Correlation (FNC) is then calculated as the correlation of the projected predicted rankings. Using this procedure, we reduce the linear dependencies of predictions on the risky features.
%This procedure can be applied to any ML prediction model
%For all models, feature projection reduces the correlation as the linear feature exposure of the prediction is removed.

%%%%%%%%%%
%%%%%%%%%%
%% Performance 
\begin{table}[htb!]
%\begin{tabular}{|l|l|l|l|l|l|}
(a) All regimes 
\\
\begin{center}
\resizebox{.9\linewidth}{!}{
\begin{tabular}{|r|l|r|r|r||r|r|}
\hline
Method                    & Feature Projection & Mean & Volatility & Max Draw & Sharpe & Calmar \\ \hline
\multirow{2}{*}{LightGBM-dart without FE}      & Yes            & 0.0215                    & 0.0182                          & 0.1153                        & \textbf{1.1806}                      & \textbf{0.1865}                      \\ \cline{2-7} 
                          & No             & 0.0278                    & 0.0284                          & 0.1622                        & 0.9791                      & 0.1714                      \\ \hline
\multirow{2}{*}{LightGBM-gbdt without FE}      & Yes            & 0.0204                    & 0.0211                          & 0.1998                        & 0.9665                      & 0.1021                      \\ \cline{2-7} 
                          & No             & 0.0262                    & 0.0321                          & 0.2378                        & 0.8140                      & 0.1102                      \\ \hline
\multirow{2}{*}{MLP without FE}      & Yes            & 0.0179                    & 0.0203                          & 0.2606                        & 0.8798                      & 0.0687                      \\ \cline{2-7} 
                          & No             & 0.0258                    & 0.0289                          & 0.1668                        & 0.8931                      & 0.1547                      \\ \hline
\end{tabular}
}
\end{center}
\bigskip
(b) High Volatility
\\
\begin{center}
\resizebox{.9\linewidth}{!}{
\begin{tabular}{|r|l|r|r|r||r|r|}
\hline
Method                    & Feature Projection & Mean & Volatility & Max Draw & Sharpe & Calmar \\ \hline
\multirow{2}{*}{LightGBM-dart without FE} & Yes            & 0.0227                    & 0.0163                          & 0.0223                        & \textbf{1.3888}                      & \textbf{1.0179}                      \\ \cline{2-7} 
                          & No             & 0.0314                    & 0.0251                          & 0.0657                        & 1.2510                      & 0.4779                      \\ \hline
\multirow{2}{*}{LightGBM-gbdt without FE} & Yes            & 0.0217                    & 0.0198                          & 0.0364                        & 1.0953                      & 0.5962                      \\ \cline{2-7} 
                          & No             & 0.0308                    & 0.0293                          & 0.1123                        & 1.0497                      & 0.2743                      \\ \hline
\multirow{2}{*}{MLP without FE} & Yes            & 0.0196                    & 0.0193                          & 0.0326                        & 1.0191                      & 0.6012                      \\ \cline{2-7} 
                          & No             & 0.0298                    & 0.0276                          & 0.1247                        & 1.0802                      & 0.2390                      \\ \hline
\end{tabular}
}
\end{center}

\bigskip
(c) Low Volatility
\\
\begin{center}
\resizebox{.9\linewidth}{!}{
\begin{tabular}{|r|l|r|r|r||r|r|}
\hline
Method                    & Feature Projection & Mean & Volatility & Max Draw & Sharpe & Calmar \\ \hline
\multirow{2}{*}{LightGBM-dart without FE}  & Yes            & 0.0206                    & 0.0195                          & 0.1153                        & \textbf{1.0576}                      & \textbf{0.1787}                      \\ \cline{2-7} 
                          & No             & 0.0252                    & 0.0305                          & 0.1622                        & 0.8257                      & 0.1554                      \\ \hline
\multirow{2}{*}{LightGBM-gbdt without FE}  & Yes            & 0.0194                    & 0.0220                          & 0.1998                        & 0.8820                      & 0.0971                      \\ \cline{2-7} 
                          & No             & 0.0227                    & 0.0338                          & 0.2378                        & 0.6727                      & 0.0955                      \\ \hline
\multirow{2}{*}{MLP without FE}  & Yes            & 0.0165                    & 0.0210                          & 0.2606                        & 0.7875                      & 0.0633                      \\ \cline{2-7} 
                          & No             & 0.0228                    & 0.0296                          & 0.1668                        & 0.7721                      & 0.1367                      \\ \hline
\end{tabular}
}
\end{center}
\caption{\textbf{The effect of standard feature projection.} Performance of different ML methods on the Numerai dataset over the test period (2015-05-15 to 2022-09-23) with and without feature projection under different market regimes: the whole test period (All), high volatility regime (High Vol), and low volatility regime (Low Vol). } 
\label{table:numerai-v4-benchmark-test-FNC}
\end{table}

%%%%%%%%%%
%%%%%%%%%%

%% Performance of models with/without projection 
In Table~\ref{table:numerai-v4-benchmark-test-FNC}, we compare the performance of the LightGBM-dart, LightGBM-gbdt and MLP with and without feature projection under different market regimes (all, high volatility, low volatility). The projection strength $\beta$ is set to 1 throughout.  
As expected, we find that the Volatility and Maximum Drawdown consistently reduced by feature projection across all models, suggesting an overall reduction of risk. 
Further, whereas feature projection improves the Sharpe and Calmar ratios of LightGBM-dart and LightGBM-gbdt under different market regimes, it does not improve the performance of MLP models consistently. 

However, the feature projection procedure suggested by Numerai based on projection out a list of `risky' features  is not optimal. In Section~\ref{section:online-learning}, we will show how online learning techniques can be used to improve the procedure. 

% Skip initial trees
\subsection{Pruning initial trees in Gradient Boosting models} 

For gradient-boosting tree models, there is a procedure related to feature projection, which can be used to reduce the excessive dependence on particular features. The procedure consists of pruning initial trees during prediction to reduce feature dependencies. Specifically, we perform a grid search over the number of initial trees to be pruned off in the trained LightGBM models, capping the number of pruned trees to not more than half of the trees to ensure our models do not degenerate.

%% Skip Initial Trees Performance 
\begin{table}[htb!]
(a) LightGBM-dart without FE 
\\
\begin{center}
\resizebox{.8\linewidth}{!}{
\begin{tabular}{|l|r|r|r||r|r|}
\hline
%Method                   & 
Pruned Trees & Mean & Volatility & Max Draw & Sharpe & Calmar \\ \hline
% LightGBM-dart without FE & 
 0          & 0.0278                    & 0.0284                          & 0.1622                        & 0.9791                      & 0.1714                      \\ \hline
%LightGBM-dart without FE & 
100        & 0.0272                    & 0.0264                          & 0.1384                        & 1.0293                      & 0.1965                      \\ \hline
%LightGBM-dart without FE & 
250        & 0.0264                    & 0.0255                          & 0.1299                        & 1.0336                      & 0.2032                      \\ \hline
%LightGBM-dart without FE & 
500      &     0.0249	 &  0.0238     &  0.1166     &     1.0459	         &         0.2136  \\ \hline
\end{tabular}
}
\end{center}
\bigskip
(b) LightGBM-gbdt without FE
\\
\begin{center}
\resizebox{.8\linewidth}{!}{
\begin{tabular}{|l|r|r|r||r|r|}
\hline
%Method                   & 
Pruned Trees & Mean & Volatility & Max Draw & Sharpe & Calmar \\ \hline
%LightGBM-gbdt without FE & 
0          & 0.0262                    & 0.0321                          & 0.2378                        & 0.8140                      & 0.1102                      \\ \hline
%LightGBM-gbdt without FE & 
100        & 0.0265                    & 0.0291                          & 0.1835                        & 0.9106                      & 0.1444                      \\ \hline
%LightGBM-gbdt without FE & 
250        & 0.0253                    & 0.0259                          & 0.1490                        & 0.9769                      & 0.1698                      \\ \hline
%LightGBM-gbdt without FE & 
500        & 0.0253                    & 0.0259                          & 0.1490                        & 0.9765                      & 0.1698                      \\ \hline
\end{tabular}
}
\end{center}
\caption{ \textbf{The effect of tree pruning.} Performance of LightGBM models in the test period (2015-05-15 to 2022-09-23, i.e., All regimes) when pruning an increasing number of initial trees.  }
\label{table:numerai-v4-benchmark-test-trees}
\end{table}

Table~\ref{table:numerai-v4-benchmark-test-trees} compares the performance of the GBDT models (LightGBM-dart and LightGBM-gbdt) under pruning of different numbers of initial trees before feature projection. Pruning initial trees during prediction improves the Sharpe and Calmar ratios of both GBDT models, but LightGBM-gbdt models see a bigger improvement than LightGBM-dart models. This is expected as LightGBM-dart models already employ a similar idea during training, i.e., the trained trees in LightGBM-dart models are already optimised. 
%Note that although the Sharpe and Calmar ratios improve with pruning, pruning 100 initial trees achieved the best mean \textbf{Corr} for both LightGBM models. This suggests 
Our numerical results also suggest that there is a natural upper limit to the trees to be pruned: there is little improvement in model performance if we prune above a threshold of 100-250 trees.

\subsection{Joint effect of feature projection and tree pruning}
%%% Combining pruning Trees and feature projection 
We then considered the joint effect of feature projection and pruning initial trees. Table \ref{table:numerai-v4-benchmark-test-trees-FNC} compares the performance (FNC) of LightGBM-dart and LightGBM-gbdt models pruning a different number of initial trees \emph{after feature projection}. The effect of pruning on model performance for both LightGBM models after feature projection is at best modest. As FNC is a measure of the effect of non-linear relationships, this suggests that in gradient boosting models, early weak learners (trees) mostly capture linear relationships whereas most of the non-linear relationships are captured in the later weak learners (trees). Therefore, pruning initial trees can be thought of as a model-dependent feature projection method.

%% Skip Initial Trees Performance after feature projection  
\begin{table}[htb!]
(a) LightGBM-dart without FE with Feature Projection
\\
\begin{center}
\resizebox{.8\linewidth}{!}{
\begin{tabular}{|l|r|r|r||r|r|}
\hline
%Method   & 
Pruned Trees & Mean & Volatility & Max Draw & Sharpe & Calmar \\ \hline
%LightGBM-dart without FE & 
0                               & 0.0215                    & 0.0182                          & 0.1153                        & 1.1806                      & 0.1865                      \\ \hline
%LightGBM-dart without FE & 
100                             & 0.0208                    & 0.0174                          & 0.1079                        & 1.1998                      & 0.1928                      \\ \hline
%LightGBM-dart without FE & 
250                             & 0.0200                    & 0.0168                          & 0.1103                        & 1.1918                      & 0.1813                      \\ \hline
%LightGBM-dart without FE & 
500                             & 0.0183                    & 0.0156                          & 0.1044                        & 1.1748                      & 0.1753                      \\ \hline
\end{tabular}
}
\end{center}
\bigskip
(b) LightGBM-gbdt without FE with Feature Projection \\
\begin{center}
\resizebox{.8\linewidth}{!}{
\begin{tabular}{|l|r|r|r||r|r|}
\hline
%Method   & 
Pruned Trees & Mean & Volatility & Max Draw & Sharpe & Calmar \\ \hline
%LightGBM-gbdt without FE & 
0                               & 0.0204                    & 0.0211                          & 0.1998                        & 0.9665                      & 0.1021                      \\ \hline
%LightGBM-gbdt without FE & 
100                             & 0.0206                    & 0.0200                          & 0.1912                        & 1.0293                      & 0.1077                      \\ \hline
%LightGBM-gbdt without FE & 
250                             & 0.0194                    & 0.0188                          & 0.2058                        & 1.0307                      & 0.0943                      \\ \hline
%LightGBM-gbdt without FE & 
500                             & 0.0193                    & 0.0188                          & 0.2063                        & 1.0301                      & 0.0936                      \\ \hline
\end{tabular}
}
\end{center}
\caption{\textbf{The joint effect of feature projection and tree pruning.} Performance of the LightGBM models in the test period (2015-05-15 to 2022-09-23, All regimes) when pruning an increasing number of initial trees after feature projection.   }
\label{table:numerai-v4-benchmark-test-trees-FNC}
\end{table}

\section{Online Learning techniques to improve post-prediction processing} 
\label{section:online-learning} 

%% What are the online learning improvements 
As a further improvement to the ML pipeline, we apply online learning approaches in two ways to deal with regime changes and non-stationarity. First, we introduce a version of feature projection called \textit{dynamic feature projection}, which applies statistical rules to determine different subsets of features to project predictions in each era. 
Second, we introduce \textit{dynamic model selection} to generate model ensembles by updating regularly the choice of model(s) from a model ensemble based on recent model performance. 

%% Background of online learning 
%\paragraph{Online learning as solving an optimal control problem}

The aim of online learning is to derive an optimal procedure to select ML models and parameters as data arrives continuously. In a continuous-time setting, the Hamilton-Jacobi-Bellman (HJB) equation is solved to find the optimal deterministic control for the decision problem \autocite{Kirk70}. The discrete-time equivalent, the Bellman equation, is used in reinforcement learning to derive optimal policies of agents~\autocite{Sutton18}. 

For the Numerai tournament, we consider online learning in the discrete-time setting, since data and predictions are required once per week. For each week $t$ ($1 \leq t \leq T$), we have a state (data) process $X_t$ that contains all the information we know about the environment (Numerai datasets and trained ML model parameters) up to week $t$. Our task is then to derive a deterministic decision process $D_t(\beta_t)$ %over time t $1 \leq t \leq T$, 
described by parameters $\beta_t(X_t)$, subject to the objective function $V_T = \max_{D_t}  \sum_{t=1}^T q(X_t,D_t)$,  where $q(X_t,D_t)$ represents the utility at time $t$ given the data and decision process. 
%This formulation can be considered as a much-simplified version of the HJB equation (without bequest value and over discrete time). 

Reinforcement Learning (RL) algorithms are commonly used to solve online learning problems. However, they are not used here due to the following reasons:
\begin{enumerate}
    \item \textit{Limited data:} Available data is not enough to train RL models, such as Deep Q Networks (DQN) \autocite{Mnih2015}, Proximal Policy Optimisation (PPO) \autocite{Schulman17} and Soft Actor-Critic (SAC) \autocite{pmlr-v80-haarnoja18b}). Generating a large number of samples is difficult here since we must avoid look-ahead bias. 

    \item \textit{Expanding action space:} Most RL algorithms~(see Ray-RLlib \autocite{Moritz17}) do not adapt naturally to an expanding action space, as is the case here. 
    %Possible workaround solutions would include creating place-holder actions which return the default actions when choosing before these actions become available to the agent. However, this induces artificial limits on the size of the action space. 
    Indeed, for dynamic model selection, the number of potential models is unbounded, as newer models trained with the latest data can be added to the candidate list. 
    %Rule-based models, on the other hand, can handle the issue of expanding action space easily.  

    \item \textit{Actions have negligible impact on environment:} Successful RL algorithms usually target applications (e.g., robotics and games \autocite{Schulman15Philipp}) where agent actions modify the environment, whereas 
    %A major advantage of reinforcement learning over other deep learning approaches is the flexibility to model complicated interactions between agent and environment. However, 
    for the trading models considered here, the trading activities have negligible market impact. 
    %namely, our action will not change the environment. 
    RL algorithms thus reduce to an online learning prediction problem in our case.
    %where the models need to generate predictions based on limited data available to the model at the time when predictions need to be made.  
    
    \item \textit{Large, correlated feature sets for projection:} 
    %In the post-prediction processing step of the pipeline, we 
    %To improve feature projection, we use a different subset of features to project predictions in each era. Yet 
    The size of the set of risky features (420 features) and their high correlation makes it computationally infeasible to learn feature subsets %Therefore it is not possible to use 
    through supervised ML 
    %to learn the optimal subset of features to project. Similarly, it is not possible 
    or RL methods, as it is difficult to construct a robust reward function. 
%    Replacing correlated features in the subset of features to project will have a limited impact on the reward function, leading to convergence issues of the reinforcement learning algorithm. Given the very complicated and weak relationships between features to project and model performance, 
    \item \textit{Model ensembling can be simplified in our setting:} The model ensemble step of the pipeline assigns portfolio weightings to different ML models. Although similar to a multi-armed bandit problem, in our case
    %However, there is a major difference between the two problems. In the multi-armed bandit problem, the probability distribution of reward from a decision is not known to the agent. 
no exploration is needed for the agent to learn the distribution of rewards from different choices
    %. In the model ensemble problem, apart from a warm-up period to build up a history of model performances, 
    since the performance of all ML models up to the decision time are known to the tournament participant. 
    %even if these models are not used. With a more complete history, there 
%    Hence there is less need to employ trial-and-error as in multi-armed bandit algorithms. 
\end{enumerate}
As a consequence, instead of reinforcement learning algorithms, we use heuristics which are shown to be effective in improving the robustness of the ML pipeline. These heuristics can be interpreted as strong priors in Bayesian learning that greatly simplify our problem. 

%    Heuristic methods thus provide suitable alternatives to learn interpretable and robust feature projection schemes. 
% Rule-based models, on the other hand, can handle the issue of expanding action space easily. 

% Dynamic feature projection
\subsection{Dynamic Feature Projection} 

In Section~\ref{sec:standard_feature_projection}, we presented the results of projecting out a subset of `risky features', which was given by Numerai and fixed throughout the whole validation and test periods. However, since market conditions are variable, we suggest changing the set of features to project in each era to adapt our ML models without the need for expensive re-training of models.
Specifically, each week we update the set of features to project based on rolling statistical properties of features, as follows. For each feature in the dataset, we calculate the correlation of the feature with the target (\textbf{Corr}) and compute lagged moving average statistics with a lag of 6 weeks to account for the lagged reporting of future performance. The look-back window to compute statistical properties of feature \textbf{Corr} is 52 weeks. We consider 5 different criteria to select the subset of features to be projected out: 
\begin{enumerate}
    \item `Fixed': 420 features provided by the portfolio optimiser in Numerai (see Section~\ref{sec:standard_feature_projection}).
    \item `Low Mean': 420 features that are least correlated to the target recently
    \item `High Mean': 420 features that are most correlated to the target recently
    \item `Low Volatility': 420 features that have correlations with lowest variability recently
    \item `High Volatility': 420 features that have correlations with largest variability recently
\end{enumerate}

%% Performance 
%% LightGBM-dart 100 models LightGBM-gbdt 0 models LightGBM-goss 0 models 
%% 
\begin{table}[htb!]
(a) LightGBM-dart without FE
\\
\begin{center}
\resizebox{.8\linewidth}{!}{
\begin{tabular}{|l|r|r|r||r|r|}
\hline
%Method                   & 
Dynamic Feature Projection & Mean & Volatility & Max Draw & Sharpe & Calmar \\ \hline
%LightGBM-dart without FE & 
Fixed                  & 0.0215                    & 0.0182                          & 0.1153                        & 1.1806                      & 0.1865                      \\ \hline
%LightGBM-dart without FE & 
Low Mean               & 0.0240                    & 0.0164                          & 0.0350                        & \textbf{1.4595}                      & \textbf{0.6857}                      \\ \hline
%LightGBM-dart without FE & 
High Mean              & 0.0218                    & 0.0185                          & 0.0986                        & 1.1783                      & 0.2211                      \\ \hline
%LightGBM-dart without FE & 
Low Vol                & 0.0244                    & 0.0200                          & 0.0538                        & 1.2220                      & 0.4535                      \\ \hline
%LightGBM-dart without FE & 
High Vol               & 0.0226                    & 0.0169                          & 0.0341                        & 1.3411                      & 0.6628                      \\ \hline
\end{tabular}
}
\end{center}
\bigskip
(b) LightGBM-gbdt without FE
\\
\begin{center}
\resizebox{.8\linewidth}{!}{
\begin{tabular}{|l|r|r|r||r|r|}
\hline
%Method                   & 
Dynamic Feature Projection & Mean & Volatility & Max Draw & Sharpe & Calmar \\ \hline
%LightGBM-gbdt without FE & 
Fixed                  & 0.0204                    & 0.0211                          & 0.1998                        & 0.9665                      & 0.1021                      \\ \hline
%LightGBM-gbdt without FE & 
Low Mean               & 0.0234                    & 0.0184                          & 0.0495                        & 1.2737                      & 0.4727                      \\ \hline
%LightGBM-gbdt without FE & 
High Mean              & 0.0199                    & 0.0212                          & 0.1469                        & 0.9381                      & 0.1355                      \\ \hline
%LightGBM-gbdt without FE & 
Low Vol                & 0.0224                    & 0.0228                          & 0.1852                        & 0.9797                      & 0.1210                      \\ \hline
%LightGBM-gbdt without FE & 
High Vol               & 0.0182                    & 0.1633                          & 0.0487                        & 1.1986                      & 0.4476                      \\ \hline
\end{tabular}
}
\end{center}
\bigskip
(c) MLP without FE 
\\
\begin{center}
\resizebox{.8\linewidth}{!}{
\begin{tabular}{|l|r|r|r|r|r|}
%Method                   & 
\hline
Dynamic Feature Neutral. & Mean & Volatility & Max Draw & Sharpe & Calmar \\ \hline
%MLP without FE           & 
Fixed                  & 0.0179                    & 0.0203                          & 0.2606                        & 0.8798                      & 0.0687                      \\ \hline
%MLP without FE           & 
Low Mean               & 0.0211                    & 0.0185                          & 0.0806                        & 1.1387                      & 0.2618                      \\ \hline
%MLP without FE           & 
High Mean              & 0.0186                    & 0.0201                          & 0.1283                        & 0.9256                      & 0.1450                      \\ \hline
%MLP without FE           & 
Low Vol                & 0.0206                    & 0.0215                          & 0.0878                        & 0.9598                      & 0.2346                      \\ \hline
%MLP without FE           & 
High Vol               & 0.0191                    & 0.0172                          & 0.0730                        & 1.1150                      & 0.2616                      \\ \hline
\end{tabular}
}
\end{center}
\caption{\textbf{The effect of Dynamic Feature Projection.} Performance of different ML models in the test period (2015-05-15 to 2022-09-23) with different dynamic feature projection methods. Fixed corresponds to the standard Feature Projection in Section~\ref{sec:standard_feature_projection}. }
\label{table:numerai-v4-benchmark-test-dynamic}
\end{table}

Table~\ref{table:numerai-v4-benchmark-test-dynamic} compares the performance of LightGBM-dart, LightGBM-gbdt and MLP models under each of the five dynamic feature projection schemes.  All Dynamic Feature Projection methods perform better than using a fixed set of features but the `Low Mean' projection method  has the best Sharpe and Calmar ratios for all ML models, followed by projection of `High Volatility' features. The worse performance of `High Mean' and 'Low Volatility' projections suggests that a large part of the model risks can be attributed to recently under-performing and high volatility features. 

Next we compared the performance of the different dynamic feature projection schemes under low volatility and high volatility market regimes, as defined in Section~ \ref{section:numerai-regime}. The results can be found in Tables \ref{table:numerai-v4-benchmark-test-dynamic-high-vol} and \ref{table:numerai-v4-benchmark-test-dynamic-low-vol} in the Supplementary Information. Whereas projecting out `Low Mean' features performs best in low volatility regimes, projecting out `Low Volatility' and 'High Mean' features performs best in high volatility regimes.
This suggests that in a low volatility regime, factors that are performing well recently continue to do so in the near future as the feature correlation structure is more stable. On the other hand, in a high volatility regime, the ML models after projecting out of `Low Volatility' and 'High Mean' features lead to improved performance. `Low Volatility' represents features that have a low variance and stable performance in the last 52 weeks. During volatile regimes, these features will under-perform; hence models that project out these features have improved performance when there is market stress. 

% Does `Low Volatility' features corresponds to more crowded features? 

%% How to select dynamic feature projection methods 
\subsection{Dynamic Model Selection}
\label{section:optimal-dfn}

In practice, it is not possible to know the best dynamic feature projection in advance. Therefore, we propose an online learning procedure consisting of two steps to select the dynamic feature projection. First, we have a warm-up period to collect data on model performances, during which all 5 feature projection methods (fixed, low mean, high mean, low volatility, high volatility) have equal weighting. The second step is to allocate weights to the optimal model based on recent performance according to the following criteria: 
\begin{itemize}
    \item `Average': Using all five feature projection methods with equal weighting 
    \item `Momentum': Using the feature projection method with the highest Mean \textbf{Corr} in the last 52 weeks
    \item `Sharpe': Using the feature projection method with the highest Sharpe Ratio in the last 52 weeks
    \item `Calmar': Using the feature projection method with the highest Calmar ratio in the last 52 weeks
\end{itemize}

\begin{table}[htb!]
(a) LightGBM-dart without FE
\\
\begin{center}
\resizebox{.8\linewidth}{!}{
\begin{tabular}{|l|r|r|r||r|r|}
\hline
%Method                   & 
Model Selection & Mean & Volatility & Max Draw & Sharpe & Calmar \\ \hline
%LightGBM-dart without FE & 
Average                & 0.0229                    & \textbf{0.0160}                          & 0.0619                        & \textbf{1.4323}                      & 0.3700                      \\ \hline
%LightGBM-dart without FE & 
Momentum               & \textbf{0.0246}                     & 0.0180                          & 0.0533                        & 1.3654                      & 0.4615                      \\ \hline
%LightGBM-dart without FE & 
Sharpe                 & 0.0234                    & 0.0165                          & 0.0533                        & 1.4148                      & 0.4390                      \\ \hline
%LightGBM-dart without FE & 
Calmar                 & 0.0225                    & 0.0171                          & \textbf{0.0350}                        & 1.3122                      & \textbf{0.6429}                      \\ \hline
\end{tabular}
}
\end{center}
\bigskip
(b) LightGBM-gbdt without FE
\\
\begin{center}
\resizebox{.8\linewidth}{!}{
\begin{tabular}{|l|r|r|r||r|r|}
\hline
%Method                   & 
Model Selection & Mean & Volatility & Max Draw & Sharpe & Calmar \\ \hline
%LightGBM-gbdt without FE & 
Average                & 0.0216                    & 0.0177                          & 0.0710                        & 1.2165                      & 0.3042                      \\ \hline
%LightGBM-gbdt without FE & 
Momentum               & 0.0228                    & 0.0201                          & 0.0729                        & 1.1342                      & 0.3128                      \\ \hline
%LightGBM-gbdt without FE & 
Sharpe                 & 0.0224                    & 0.0187                          & 0.0729                        & 1.1966                      & 0.3073                      \\ \hline
%LightGBM-gbdt without FE & 
Calmar                 & 0.0216                    & 0.0195                          & 0.0508                        & 1.1102                      & 0.4252                      \\ \hline
\end{tabular}
}
\end{center}
\bigskip
(c) MLP without FE \\
\begin{center}
\resizebox{.8\linewidth}{!}{
\begin{tabular}{|l|r|r|r||r|r|}
%Method                   & 
\hline
%Method                   & 
Model Selection & Mean & Volatility & Max Draw & Sharpe & Calmar \\ 
\hline
%MLP without FE           & 
Average                & 0.0195                    & 0.0175                          & 0.0918                        & 1.1149                      & 0.2124                      \\ \hline
%MLP without FE           & 
Momentum               & 0.0212                    & 0.0191                          & 0.0878                        & 1.1124                      & 0.2415                      \\ \hline
%MLP without FE           & 
Sharpe                 & 0.0207                    & 0.0186                          & 0.0878                        & 1.1110                      & 0.2358                      \\ \hline
%MLP without FE           & 
Calmar                 & 0.0187                    & 0.0201                          & 0.1973                        & 0.9309                      & 0.0948                      \\ \hline
\end{tabular}
}
\end{center}
\caption{\textbf{The effect of dynamic model selection.} Performance of different ML models in the test period (2015-05-15 to 2022-09-23) with different online learning procedures selecting the optimal dynamic feature projection method. }
\label{table:numerai-v4-benchmark-test-dynamic-FN-optimal}
\end{table}

In Table \ref{table:numerai-v4-benchmark-test-dynamic-FN-optimal}, we use these criteria to select the optimal dynamic feature projection based on recent performance. As above, a lag of 6 weeks is applied to account for data delays. We find that for all three ML models (LightGBM-dart/LightGBM-gbdt/MLP), the `Momentum' selection method has higher 'Mean' \text{Corr} and Calmar ratio than the`Average' and `Sharpe' methods. This shows that the `Momentum' method, a very simple model selection method that chooses the recent best-performing model, can adapt a trained ML model towards different market regimes efficiently. However, we find that the simple 'Average' scheme has the lowest 'Volatility' and highest Sharpe ratio. 
% Calmar 
As expected, the `Calmar' selection method gives a lower Maximum Drawdown and higher Calmar ratio than all the other dynamical model selection algorithms. However, for MLP models, the `Calmar' selection method significantly underperforms, with a much higher Max Drawdown. This suggests that selection based on historical drawdown is not robust, especially under situations with regime changes. 

In summary, the proposed online learning procedure to select optimal dynamic feature engineering methods can significantly reduce trading risks and improve the robustness of trading models by considering simple schemes to select or average the different dynamic feature projection methods.

\section{Discussion} 

\paragraph{Summary:}
%% Benchmark 
Motivated by the Numerai tournament, we have designed an ML pipeline that can be applied to tabular temporal data of stock prices to underpin strategies for trading of market-neutral stock portfolios. The various steps in the ML pipeline are designed for robustness against regime changes and to avoid information leakage through time by using models with relatively low complexity, so as to reduce the danger of overfitting, and with high robustness to changes in hyperparameters and other choices in the algorithms. 

%% Choice of ML models 
Regarding the choice of ML models, we find that gradient-boosting decision tree models are both more robust and interpretable than neural network-based models, and they allow more consistent performance under different market regimes. 

%% Post-prediction processing
We also find that post-prediction processing using online learning techniques, which are model-agnostic, is an effective means of adapting trained ML models towards new situations without the need to retrain ML models and introduce additional model uncertainty.  Firstly, using dynamic feature projection produces models with different flavours in an interpretable way, which also have better risk-adjusted performance than models with fixed feature projection. 
%% Model Ensemble 
Secondly, dynamic model selection can be applied  in guiding the selection of an optimal model(s) from a growing model ensemble. We find that a simple design, such as equal-weighted models, has a robust performance under different market regimes, but selecting the best model based on recent performance can provided an improvement as it switches to a lower-risk model during more volatile market regimes. 
%

%% Robustness of Methods 
\paragraph{Robustness of the pipeline:}
An important consideration in this work is to provide a methodology that is robust to diverse sources of variability. Although not presented in the main text, in  Section \ref{section:robustness} of the Supplementary Information, we have carried out an extensive study of the robustness of our ML pipeline under three different sources of variability: (i)  different random seeds in the training of algorithms; (ii)  changes in data splits for cross-validation; and (iii) randomness in feature selection for feature projection. 
The results show that LightGBM dart models are robust against these changes. The statistical rules used in dynamic feature projection are also shown to perform better than features chosen at random.

%% Further Work 
\paragraph{Further work:} Finally, we discuss some ideas for further work to improve the ML pipeline presented here. 

Staking is commonly used in ML competitions to improve the robustness of models. The dynamic model selection method suggested in this study falls into this broad category, whereby different models can be chosen based on certain statistical criteria that mirror different strategies under regime changes. 
It remains an open research area into how reinforcement learning or other online learning methods can be used to learn optimal staking weights between different ML models, given their historical performance and correlations. 

%%% Measuring model contribution 
The diversity of models within a model ensemble is a key ingredient for dynamic model selection and other model ensemble/staking methods. A new metric could be designed to study the impact of a new ML model on an existing model ensemble. This metric could then be used to train new ML models that are uncorrelated to existing ones.  

%%% Feature Engineering 
We also note that although simple feature engineering methods did not improve performance here (see Table~\ref{table:numerai-v4-benchmark}), an open direction could be to identify robust relationships between features over different regimes using generative models (e.g., Variational Autoencoders \autocite{DBLP:journals/corr/KingmaW13}) to create new features that could capture meaningful and non-trivial non-linear relationships.

%% Conclusion 
\paragraph{Conclusion:}
Overall, our results suggest using simple, well-established ML models, such as gradient-boosting decision trees, instead of specialised neural network models for this task. Rather than using a single neural network to perform feature engineering, model training/inference and post-prediction transformations, our modularised design of the ML pipeline offers increased robustness and transparency. Researchers can add, modify or delete a component without affecting the rest of the pipeline. Creating model ensembles improves model performance by reducing idiosyncratic variance from individual ML models. The simple model selection rules based on recent performance provide a baseline that works well under different market regimes, whereas various portfolio metrics such as Sharpe and Calmar ratios are improved by using the recently best-performing models.

%% Computational Resources Constraint / Distributed Learning 
The Gradient Boosting models used here are suitable for distributed learning, where large datasets are split into smaller batches to train on different machines with varying computational limitations. Data science competitions like the Numerai tournament rely on community efforts of individual data scientists to create a meta-model. Such crowdsourcing relies on the assumption that a complicated ML model, which would need to be trained with advanced hardware, can be approximated by combining a number of simpler ML models (each trained with fewer data or features). Studying the convergence of model performance would be important for organising the data science competition as it decides how many participants are needed to maintain a well-diversified pool of models to create the meta-model.

% Update Link once migrated to Group Repo 
\section{Data and Code Availability} 
The data and code used in this paper are available at \url{https://github.com/barahona-research-group/THOR-1}.

\section{Acknowledgement} 
This work was supported in part by the Wellcome Trust under Grant 108908/B/15/Z and by the EPSRC under grant EP/N014529/1 funding the EPSRC Centre for Mathematics of Precision Healthcare at Imperial. MB also acknowledges support by the Nuffield Foundation under the project ``The Future of Work and Well-being: The Pissarides Review". We thank Numerai GP, LLC for providing the datasets used in the study.

%Bibliography
%\bibliographystyle{apacite} 
\printbibliography
%http://www.ctan.org/tex-archive/biblio/bibtex/contrib/apacite/apacite.pdf

\newpage 

\section{Supplementary Information}

\subsection{Dynamic Feature Projection for low and high volatility regimes} 
%Here we show the performance of dynamic feature projection for low and high volatility regimes.

\begin{table}[H]
\textbf{Low Volatility}\\~\\
(a) LightGBM-dart without FE
\\
\begin{center}
\resizebox{.8\linewidth}{!}{
\begin{tabular}{|l|r|r|r||r|r|}
\hline
Feature Projection & \multicolumn{1}{l|}{Mean} & \multicolumn{1}{l|}{Volatility} & \multicolumn{1}{l|}{Max Draw} & \multicolumn{1}{l|}{Sharpe} & \multicolumn{1}{l|}{Calmar} \\ \hline
Fixed                  & 0.0206                    & 0.0195                          & 0.1153                        & 1.0576                      & 0.1787                      \\ \hline
Low Mean               & 0.0255                    & 0.0175                          & 0.0350                        & \textbf{1.4578}                      & \textbf{0.7286}                      \\ \hline
 High Mean              & 0.0207                    & 0.0206                          & 0.0986                        & 1.0033                      & 0.2099                      \\ \hline
 Low Vol                & 0.0238                    & 0.0221                          & 0.0538                        & 1.0793                      & 0.4424                      \\ \hline
 High Vol               & 0.0235                    & 0.0180                          & 0.0341                        & 1.3069                      & 0.6891                      \\ \hline
\end{tabular}
}
\end{center}
\bigskip
(b) LightGBM-gbdt without FE
\\
\begin{center}
\resizebox{.8\linewidth}{!}{
\begin{tabular}{|l|r|r|r||r|r|}
\hline
Feature Projection & \multicolumn{1}{l|}{Mean} & \multicolumn{1}{l|}{Volatility} & \multicolumn{1}{l|}{Max Draw} & \multicolumn{1}{l|}{Sharpe} & \multicolumn{1}{l|}{Calmar} \\ \hline
Fixed                  & 0.0194                    & 0.0220                          & 0.1998                        & 0.8820                      & 0.0971                      \\ \hline
Low Mean               & 0.0251                    & 0.0188                          & 0.0495                        & 1.3328                      & 0.5071                      \\ \hline
High Mean              & 0.0184                    & 0.0228                          & 0.1469                        & 0.8053                      & 0.1253                      \\ \hline
Low Vol                & 0.0214                    & 0.0247                          & 0.1852                        & 0.8657                      & 0.1150                       \\ \hline
High Vol               & 0.0225                    & 0.0188                          & 0.0487                        & 1.1939                      & 0.4620                      \\ \hline
\end{tabular}
}
\end{center}
\bigskip
(c) MLP without FE
\\
\begin{center}
\resizebox{.8\linewidth}{!}{
\begin{tabular}{|l|r|r|r||r|r|}
\hline
Feature Projection & \multicolumn{1}{l|}{Mean} & \multicolumn{1}{l|}{Volatility} & \multicolumn{1}{l|}{Max Draw} & \multicolumn{1}{l|}{Sharpe} & \multicolumn{1}{l|}{Calmar} \\ \hline
Fixed                  & 0.0165                    & 0.0210                          & 0.2606                        & 0.7875                      & 0.0633                      \\ \hline
Low Mean               & 0.0215                    & 0.0187                          & 0.0496                        & 1.1496                      & 0.4335                      \\ \hline
High Mean              & 0.0170                    & 0.0210                          & 0.1283                        & 0.8118                      & 0.1325                      \\ \hline
Low Vol                & 0.0194                    & 0.0229                          & 0.0878                        & 0.8487                      & 0.2210                      \\ \hline
High Vol               & 0.0194                    & 0.0177                          & 0.0730                        & 1.0990                      & 0.2658                      \\ \hline
\end{tabular}
}
\end{center}
\caption{Performance of ML models in the test period (2015-05-15 to 2022-09-23) with different dynamic feature projection methods in low volatility regime }
\label{table:numerai-v4-benchmark-test-dynamic-low-vol}
\end{table}

\newpage

\begin{table}[H]
\textbf{High Volatility}\\~\\
(a) LightGBM-dart without FE
\\
\begin{center}
\resizebox{.8\linewidth}{!}{
\begin{tabular}{|l|r|r|r||r|r|}
\hline
Feature Projection & \multicolumn{1}{l|}{Mean} & \multicolumn{1}{l|}{Volatility} & \multicolumn{1}{l|}{Max Draw} & \multicolumn{1}{l|}{Sharpe} & \multicolumn{1}{l|}{Calmar} \\ \hline
Fixed                  & 0.0227                    & 0.0163                          & 0.0223                        & 1.3888                      & 1.0179                      \\ \hline
Low Mean               & 0.0220                    & 0.0148                          & 0.0199                        & 1.4907                      & 1.1055                      \\ \hline
High Mean              & 0.0233                    & 0.0151                          & 0.0206                        & \textbf{1.5372}                      & \textbf{1.1311}                      \\ \hline
Low Vol                & 0.0252                    & 0.0168                          & 0.0330                        & 1.4980                      & 0.7636                      \\ \hline
High Vol               & 0.0215                    & 0.0152                          & 0.0143                        & 1.4077                      & 1.5035                      \\ \hline
\end{tabular}
}
\end{center}
\bigskip
(b) LightGBM-gbdt without FE
\\
\begin{center}
\resizebox{.8\linewidth}{!}{
\begin{tabular}{|l|r|r|r||r|r|}
\hline
Feature Projection & \multicolumn{1}{l|}{Mean} & \multicolumn{1}{l|}{Volatility} & \multicolumn{1}{l|}{Max Draw} & \multicolumn{1}{l|}{Sharpe} & \multicolumn{1}{l|}{Calmar} \\ \hline
 Fixed                  & 0.0217                    & 0.0198                          & 0.0364                        & 1.0953                      & 0.5962                      \\ \hline
Low Mean               & 0.0212                    & 0.0176                          & 0.0380                        & 1.2039                      & 0.5579                      \\ \hline
High Mean              & 0.0218                    & 0.0186                          & 0.0334                        & 1.1728                      & 0.6527                      \\ \hline
Low Vol                & 0.0237                    & 0.0201                          & 0.0306                        & 1.1792                      & 0.7745                      \\ \hline
High Vol               & 0.0209                    & 0.0173                          & 0.0308                        & 1.2068                      & 0.6786                      \\ \hline
\end{tabular}
}
\end{center}
\bigskip
(c) MLP without FE
\\
\begin{center}
\resizebox{.8\linewidth}{!}{
\begin{tabular}{|l|r|r|r||r|r|}
\hline
Feature Projection & \multicolumn{1}{l|}{Mean} & \multicolumn{1}{l|}{Volatility} & \multicolumn{1}{l|}{Max Draw} & \multicolumn{1}{l|}{Sharpe} & \multicolumn{1}{l|}{Calmar} \\ \hline
Fixed                  & 0.0196                    & 0.0193                          & 0.0326                        & 1.0191                      & 0.6012                      \\ \hline
Low Mean               & 0.0205                    & 0.0183                          & 0.0806                        & 1.1212                      & 0.2543                      \\ \hline
High Mean              & 0.0170                    & 0.0210                          & 0.1283                        & 0.8118                      & 0.1325                      \\ \hline
Low Vol                & 0.0222                    & 0.0194                          & 0.0397                        & 1.1442                      & 0.5592                      \\ \hline
High Vol               & 0.0187                    & 0.0165                          & 0.0336                        & 1.1368                      & 0.5565                      \\ \hline
\end{tabular}
}
\end{center}
\caption{Performance of ML models in the test period (2015-05-15 to 2022-09-23) with different dynamic feature projection methods in high volatility regime }
\label{table:numerai-v4-benchmark-test-dynamic-high-vol}
\end{table}
\pagebreak

%%% hyperparameter Space for ML models 
\newpage
\subsection{Hyperparameter search space for different ML models}
\label{section:optuna}

We ran all experiments on a GPU cluster, each node of which contains a NVIDIA GeForce RTX 2080 Ti GPU, running with 4352 CUDA cores and 11GB memory. Hyperparameter search is performed using Optuna \autocite{Takuya19}. For each Feature Engineering/ML pipeline, hyperparameter search is ran for at most 8 hours or at most 100 configurations, whichever came first. The default TPE sampler in Optuna is used to perform the hyperparameter search. 

In Figure \ref{numerai-hyperspace-1} and \ref{numerai-hyperspace-2}, we list the hyperparameter search parameters defined in Optuna  \autocite{Takuya19} for different ML models used in the main text to train the models. For MLP and TabNet, we use Adam \autocite{https://doi.org/10.48550/arxiv.1412.6980} as the gradient optimiser, using the recommended learning rates in their implementations. 

\begin{figure}[H]
\begin{itemize}
    \item Feature Engineering 
    \begin{itemize}
    \item Numerai Basic Feature Engineering 
        \begin{itemize}
            \item dropout pct: low:0.05, high:0.25, step:0.05,
            \item no product features: low:50, high:1000, step:50, 
        \end{itemize}
    \end{itemize}    
    \item ML Models 
    \begin{itemize}
        \item LightGBM-gbdt 
        \begin{itemize}
            \item n estimators: low:50, high:1000, step:50
            \item learning rate: low:0.005, high:0.1, log:True 
            \item min data in leaf: low:2500, high:40000, step:2500
            \item lambda l1: low:0.01, high: 1, log:True
            \item lambda l2: low:0.01, high: 1, log:True
            \item feature fraction: low:0.1, high:1, step:0.05
            \item bagging fraction: low:0.5, high:1, step:0.05
            \item bagging freq: low:10, high:50, step:10
        \end{itemize}
        \item LightGBM-dart 
        \begin{itemize}
            \item n estimators: low:50, high:1000, step:50
            \item learning rate: low:0.005, high:0.1, log:True 
            \item min data in leaf: low:2500, high:40000, step:2500
            \item lambda l1: low:0.01, high: 1, log:True
            \item lambda l2: low:0.01, high: 1, log:True
            \item feature fraction: low:0.1, high:1, step:0.05
            \item bagging fraction: low:0.5, high:1, step:0.05
            \item bagging freq: low:10, high:50, step:10
            \item drop rate: low:0.1, high:0.5, step:0.1
            \item skip drop: low:0.1, high:0.8, step:0.1
        \end{itemize}     
        \item LightGBM-goss
        \begin{itemize}
            \item n estimators: low:50, high:1000, step:50
            \item learning rate: low:0.005, high:0.1, log:True 
            \item min data in leaf: low:2500, high:40000, step:2500
            \item lambda l1: low:0.01, high: 1, log:True
            \item lambda l2: low:0.01, high: 1, log:True
            \item feature fraction: low:0.1, high:1, step:0.05
            \item bagging fraction: low:0.5, high:1, step:0.05
            \item bagging freq: low:10, high:50, step:10
            \item top rate: low:0.1, high:0.4, step:0.05
            \item other rate: low:0.05, high:0.2, step:0.05
        \end{itemize}          
    \end{itemize}        
\end{itemize}
\caption{Hyperparameter Space for ML models }    
\label{numerai-hyperspace-1}
\end{figure}

\begin{figure}[H]
\begin{itemize}
    \item Deep Learning models 
    \begin{itemize}
        \item MLP
        \begin{itemize}
            \item max epochs: low:10, high:100, step:5
            \item patience: low:5, high:20, step:5
            \item num layers: low:2, high:7, step:1 
            \item neurons: low:64, high:1024, step:64
            \item neuron scale: low:0.3, high:1, log:True
            \item dropout: low:0.1, high:0.9, log:True
            \item batch size: low:10240, high:40960, step:10240
        \end{itemize}        
        \item TabNet
        \begin{itemize}
            \item max epochs: low:10, high:100, step:5
            \item patience: low:5, high:20, step:5
            \item batch size: low:1024, high:4096, step:1024
            \item num d: low:4, high:16, step:4
            \item num a: low:4, high:16, step:4
            \item num steps: low:1, high:3, step:1
            \item num shared: low:1, high:3, step:1
            \item num independent: low:1, high:3, step:1
            \item gamma : low:1, high:2, step:0.1
            \item momentum: low:0.01, high:0.4, step:0.01
            \item lambda sparse: low:0.0001, high:0.01, log:True 
        \end{itemize}         
    \end{itemize}        
\end{itemize}
\caption{Hyperparameter Space for ML models }    
\label{numerai-hyperspace-2}
\end{figure}
\pagebreak

\newpage
\subsection{Robustness of ML pipeline} 
\label{section:robustness}

One of our aims is to provide a robust pipeline for tabular temporal data under regime changes. Here we present additional results checking the robustness of the method under different scenarios and sources of variability.

\paragraph{Robustness under changes of random seeds in the learning algorithms} 

In Table~\ref{table:numerai-v4-benchmark-test-random-seeds}, we report the variability of the performance of the LightGBM-dart, LightGBM-gbdt and MLP models trained starting from 10 different initial random seeds.  
The performance is generally robust to the change in random seeds, with small variances in the prediction of the mean \textbf{Corr} and volatility and moderate for the Maximum Drawdown.   
\begin{table}[H]
\begin{center}
\resizebox{.95\linewidth}{!}{
\begin{tabular}{|l|r|r|r|r||r|r|}
\hline
Model  &  & Mean & Volatility & Max Draw & Sharpe & Calmar \\ \hline 
\multirow{2}{*}{
LightGBM-dart without FE} & Mean & 0.0254  & 0.0266  & 0.1567  & 0.9593 & 0.1639   \\  \cline{2-7} 
 & S.D.   & 0.0006  & 0.0007  & 0.0158   & 0.0365 & 0.0175  \\ \hline
\multirow{2}{*}{LightGBM-gbdt without FE} & Mean  & 0.0253 & 0.0312 & 0.2338  & 0.8104 & 0.1100   \\ \cline{2-7} 
& S.D.  & 0.0006  & 0.0006  & 0.0296   & 0.0278 & 0.0153  \\ \hline
\multirow{2}{*}{
MLP without FE} & Mean & 0.0233 & 0.0271 & 0.1643 & 0.8600  & 0.1446   \\ \cline{2-7} 
& S.D.   & 0.0009  & 0.0011  & 0.0248   & 0.0365 & 0.0219  \\ \hline
\end{tabular}
}
\end{center}
\caption{Variability of the performance of ML models in the test period (2015-05-15 to 2022-09-23). The mean and standard deviation of each portfolio metrics are calculated over models with 10 different random seeds for each method }
\label{table:numerai-v4-benchmark-test-random-seeds}
\end{table}

%% How to average models 
A general strategy to reduce the variance of the predictions is to combine different ML models. There are two ways to do so: (i) averaging over models, i.e., averaging the performance of different models; and (ii) averaging over predictions, i.e., averaging predictions from each model and then scoring the average prediction from different models against the target. 
In Table \ref{table:numerai-v4-benchmark-test-average-methods}, we apply these two types of averaging to the 10 models obtained training with different initial random seeds, whose average and variance predictions are reported in Table~\ref{table:numerai-v4-benchmark-test-random-seeds} To do so, we create 6 subsets of 5 models, e.g., subset 1 contains the models with random seeds 1-5; subset 2 contains the models with random seeds 2-6; etc; and subset 6 contains the models with random seeds 6-10.  
We find that averaging over predictions from the different models gives higher mean \textbf{Corr} and Sharpe/Calmar ratios. Therefore, this averaging method is used to compute model performances in Table~\ref{table:numerai-v4-benchmark} in the main text. Model variances of the averaged predictions are also much lower than the individual models. 

\begin{table}[H]
\begin{center}
\resizebox{.95\linewidth}{!}{
\begin{tabular}{|l|l|l|l|l|l|l|l|}
\hline
Model                          & Average                           &      & Mean   & Volatility & Max Draw & Sharpe & Calmar \\ \hline
\multirow{4}{*}{LightGBM-dart without FE} & \multirow{2}{*}{Over Models}      & Mean & 0.0255 & 0.0259     & 0.1500   & 0.9832 & 0.1704 \\ \cline{3-8} 
                               &                                   & S.D. & 0.0002 & 0.0001     & 0.0081   & 0.0094 & 0.0102 \\ \cline{2-8} 
                               & \multirow{2}{*}{Over Predictions} & Mean & 0.0276 & 0.0282     & 0.1605   & 0.9765 & 0.1721 \\ \cline{3-8} 
                               &                                   & S.D. & 0.0002 & 0.0001     & 0.0087   & 0.0089 & 0.0104 \\ \hline
\multirow{4}{*}{LightGBM-gbdt without FE} & \multirow{2}{*}{Over Models}      & Mean & 0.0253 & 0.0308     & 0.2233   & 0.8196 & 0.1134 \\ \cline{3-8} 
                               &                                   & S.D. & 0.0001 & 0.0002     & 0.0080   & 0.0062 & 0.0044 \\ \cline{2-8} 
                               & \multirow{2}{*}{Over Predictions} & Mean & 0.0260 & 0.0319     & 0.2306   & 0.8161 & 0.1131 \\ \cline{3-8} 
                               &                                   & S.D. & 0.0001 & 0.0002     & 0.0083   & 0.0061 & 0.0044 \\ \hline
\multirow{4}{*}{MLP without FE}           & \multirow{2}{*}{Over Models}      & Mean & 0.0231 & 0.0259     & 0.1490   & 0.8922 & 0.1550 \\ \cline{3-8} 
                               &                                   & S.D. & 0.0003 & 0.0003     & 0.0055   & 0.0095 & 0.0065 \\ \cline{2-8} 
                               & \multirow{2}{*}{Over Predictions} & Mean & 0.0254 & 0.0284     & 0.1610   & 0.8930 & 0.1576 \\ \cline{3-8} 
                               &                                   & S.D. & 0.0002 & 0.0004     & 0.0062   & 0.0078 & 0.0050 \\ \hline
\end{tabular}

}
\end{center}
\caption{Performance of different ML methods on Numerai v4 dataset in the test period (2015-05-15 to 2022-09-23) with two different averaging methods }
\label{table:numerai-v4-benchmark-test-average-methods}
\end{table}

\paragraph{Robustness under different cross-validation data splits} 

As financial data are regime dependent, an important measure of model robustness is to measure the performance of ML models that have been trained using different cross-validation splits of the data and compute how much the model performance changes over different test periods. 

To ascertain the robustness of data splits, we have carried out 3 cross-validation splits (CV 1, CV 2, CV 3) as shown in Table~\ref{table:cross-validation-online}. The hyperparameters are optimised under CV 1, which is the cross-validation used to generate the model performances in the main text. These hyperparameters are fixed for the models trained under the CV 2 and CV 3 splits. For ML methods that require early stopping, the data in the validation period (different for each split) are used to regularise the models. Therefore, by reusing the optimised hyperparameters across all splits, we evaluate the robustness of the model performance to the optimisation of hyperparameters. We then compute the performance when applying the models to shifted cross-validation datasets in the walk-forward CV 2 and CV 3 data splits.
Our results show good consistency in performance across CV 2 and CV 3, with only a small deterioration of the results as compared to CV 1 (over which the hyperparameters were optimised). 
We also find that LightGBM-dart with FE, the ML method that has the highest mean \textbf{Corr} in CV 1, has the greatest return and best Sharpe and Calmar ratios also in other cross-validations, as seen in Table~\ref{table:numerai-v4-benchmark-walk-forward-cv}.

\begin{table}[H]
\begin{center}
\resizebox{.95\linewidth}{!}{
\begin{tabular}{|l|l|l|l|l|l|}
\hline
     & Train Start & Train End  & Validation Start & Validation End & Enter Ensemble  \\ \hline
CV 1 & 2003-01-03  & 2012-07-27 & 2012-12-21       & 2014-11-14     & 2015-05-15  \\ \hline
CV 2 & 2003-01-03  & 2014-06-27 & 2014-11-21       & 2016-10-14     & 2017-04-14    \\ \hline
CV 3 & 2003-01-03  & 2016-05-27 & 2016-10-21       & 2018-09-14     & 2019-03-15    \\ \hline
% CV 4 & 2003-01-03  & 2018-04-27 & 2018-09-21       & 2020-08-14     & 2021-02-12    \\ \hline
\end{tabular}
}
\end{center}
\caption{Various cross-validation schemes to train ML models on different parts of the data. CV 1 is the cross-validation used for hyperparameter optimisation and training ML models in the main text. }
\label{table:cross-validation-online}
\end{table}

\begin{table}[H]
(a) CV 1 (2015-05-15 to 2022-09-23) 
\\
\begin{center}
\resizebox{.8\linewidth}{!}{
\begin{tabular}{|r|r|r|r||r|r|}
\hline
Method                & Mean   & Volatility & Max Draw & Sharpe & Calmar \\ \hline
LightGBM-dart without FE   & 0.0278 & 0.0284     & 0.1622   & 0.9791 & 0.1714 \\ \hline
LightGBM-gbdt without FE   & 0.0262 & 0.0321     & 0.2378   & 0.8140 & 0.1102 \\ \hline
MLP without FE             & 0.0258 & 0.0289     & 0.1668   & 0.8931 & 0.1547 \\ \hline
\end{tabular}
}
% \label{table:numerai-v4-benchmark-validate}
%  \end{table}
\end{center}
\bigskip
(b) CV 2 (2017-04-14 to 2022-09-23)
\\
\begin{center}
\resizebox{.8\linewidth}{!}{
\begin{tabular}{|r|r|r|r||r|r|}
\hline
Method                & Mean   & Volatility & Max Draw & Sharpe & Calmar \\ \hline
LightGBM-dart without FE   & 0.0250 & 0.0278     & 0.1817   & 0.8990 & 0.1376 \\ \hline
LightGBM-gbdt without FE   & 0.0231 & 0.0324     & 0.3227   & 0.7104 & 0.0716 \\ \hline
MLP without FE             & 0.0215 & 0.0289     & 0.2307   & 0.7446 & 0.0932 \\ \hline
\end{tabular}
}
% \label{table:numerai-v4-benchmark-validate}
%  \end{table}
\end{center}

%
% \begin{table}[htb!]
\bigskip
(c) CV 3 (2019-03-15 to 2022-09-23) 
\\
\begin{center}
\resizebox{.8\linewidth}{!}{
\begin{tabular}{|r|r|r|r||r|r|}
\hline
Method                     & Mean   & Volatility & Max Draw & Sharpe & Calmar \\ \hline
LightGBM-dart without FE   & 0.0264 & 0.0297     & 0.1380   & 0.8140 & 0.1913 \\ \hline
LightGBM-gbdt without FE   & 0.0261 & 0.0336     & 0.1584   & 0.7772 & 0.1648 \\ \hline
MLP without FE             & 0.0224 & 0.0240     & 0.1171   & 0.9339 & 0.1913 \\ \hline
\end{tabular}
}
\end{center}
\caption{Performance of selected machine learning methods on the Numerai dataset in the test period for various walk-forward cross-validation schemes, (a) CV 1, (b) CV 2 and (c) CV 3  }
\label{table:numerai-v4-benchmark-walk-forward-cv}
\end{table}

\paragraph{Robustness under feature selection for dynamic feature projection} 

A fixed set of 420 features to be projected was given by the Numerai organisers based on internal evaluations of parameters.   
In Section \ref{section:online-learning}, we introduce several statistical rules that allow us to select a varying subset of features to be projected in each era based on empirical heuristic criteria motivated by financial modelling.  

To evaluate the robustness of the proposed statistical rules, we draw 100 subsets of 420 features selected at random. and use each set to project the raw predictions from ML models. We then evaluate the performance of ML models based on each of the random subsets.
Using the procedure described in section \ref{section:optimal-dfn} we then select the optimal dynamic feature projection method and compute the performance of the top 10 models of the highest mean \textbf{Corr}, Sharpe and Calmar ratio over the test period. The results are reported in Table \ref{table:numerai-v4-benchmark-test-dynamic-FN-random} and should be compared to the performance of the same models in Table~\ref{table:numerai-v4-benchmark-test-dynamic-FN-optimal}, which were obtained with dynamic feature projection using the statistical rules defined in section  \ref{section:optimal-dfn}. 

The mean \textbf{Corr} of models obtained with random feature projection for each rule (Momentum/Sharpe/Calmar) are lower than those obtained using the statistical rules in Table~\ref{table:numerai-v4-benchmark-test-dynamic-FN-optimal}. On the other hand, the Sharpe ratio of models for models with random feature projection is slightly higher, as expected due to the variance reduction effect by averaging over 10 different models. For models selected based on the Calmar rule, the models obtained with statistical rules have a much higher Calmar ratio than random feature projection. It suggests the statistical rules defined can effectively reduce model risks by reducing linear exposure to undesirable features.

\begin{table}[H]
(a) LightGBM-dart without FE
\\
\begin{center}
\resizebox{.8\linewidth}{!}{
\begin{tabular}{|l|r|r|r||r|r|}
\hline
%Method & 
Feature Projection & Mean   & Volatility & Max Draw & Sharpe & Calmar \\ \hline
Average & 0.0214  & 0.0147  & 0.0482    & 1.4547   & 0.4440  \\ \hline
Momentum  & 0.0216 & 0.0149   & 0.0472  & 1.4522  & 0.4576  \\ \hline
Sharpe  & 0.0213  & 0.0147  & 0.0459    & 1.4474  & 0.4641  \\ \hline
Calmar & 0.0214  & 0.0148  & 0.0453     & 1.4504  & 0.4724   \\ \hline
\end{tabular}
}
\end{center}
\bigskip

(b) LightGBM-gbdt without FE 
\\
\begin{center}
\resizebox{.8\linewidth}{!}{
\begin{tabular}{|l|r|r|r||r|r|}
\hline
%Method & 
Feature Projection & Mean   & Volatility & Max Draw & Sharpe & Calmar \\ \hline
Average  & 0.0203  & 0.0167 & 0.0664          & 1.2140  & 0.3057   \\ \hline
Momentum  & 0.0208  & 0.0167   & 0.0641      & 1.2457   & 0.3245   \\ \hline
 Sharpe  & 0.0206  & 0.0168  & 0.0618          & 1.2267  & 0.3333  \\ \hline
 Calmar  & 0.0216   & 0.0195   & 0.0508       & 1.1102   & 0.2743  \\ \hline
\end{tabular}
}
\end{center}
\bigskip 

(c) MLP without FE 
\\
\begin{center}
\resizebox{.8\linewidth}{!}{
\begin{tabular}{|l|r|r|r||r|r|}
\hline
%Method & 
Feature Projection & Mean   & Volatility & Max Draw & Sharpe & Calmar \\ \hline
Average  & 0.0176  & 0.0165 & 0.0831          & 1.0658  & 0.2118   \\ \hline
Momentum  & 0.0179  & 0.0165   & 0.0790      & 1.0842   & 0.2266   \\ \hline
 Sharpe  & 0.0177  & 0.0164 & 0.0762          & 1.0751  & 0.2323  \\ \hline
 Calmar  & 0.0175   & 0.0167   & 0.0825       & 1.0511   & 0.2121  \\ \hline
\end{tabular}
}
\end{center}

\caption{Performance of different ML models in the test period (2015-05-15 to 2022-09-23) obtained with random feature projection. These are averages obtained by selecting the top 10 models under the different online learning procedures over the test period. }
\label{table:numerai-v4-benchmark-test-dynamic-FN-random}
\end{table}

\newpage 

\subsection{Pseudocode for algorithms in the text}
For completeness, we present here brief pseudocode for some of the main methods in the paper with the appropriate references.

\begin{algorithm}[htb!]
Given $N$ data samples $(\mathbf{x_i}, y_i), 1 
\leq i \leq N$ with the aim to find an increasing better estimate $\hat{f}(\mathbf{x})$ of the minimising function $f(x)$ which minimise the loss $\mathcal{L}(f)$ between targets and predicted values. $\mathcal{L}(f) = \sum_i l(y_i,f(\mathbf{x_i})) $ where $l$ is a given loss function such as mean square losses for regression problems. Function $f$ is restricted to the class of additive models $f(\mathbf{x}) = \sum_{k=1}^K w_k h(\mathbf{x},	\bm{\alpha_k})$ where $h(\cdot,\bm{\alpha})$ is a weak learner with parameters $\bm{\alpha}$ and $w_k$ are the weights. \\

Initialise  $f_0(\mathbf{x}) = \arg \min_{\bm{\alpha_0}} \sum_{i=1}^N l(y_i, h(\mathbf{x_i}, 	\bm{\alpha_0}))$  \\

\For{k = 1 : K}{
    Compute the gradient residual using $g_{ik} = - \left [ \frac{\partial l(y_i, f_{k-1}(\mathbf{x_i})) }{\partial f_{k-1}(\mathbf{x_i}) } \right ]  $ \\
    Use the weak learner to compute $\bm{\alpha_k}$ which minimises $ \sum_{i=1}^N (g_{ik} - h(\mathbf{x_i},	\bm{\alpha_k}))^2 $  \\
    Update with learning rate $\lambda$ $f_k(\mathbf{x}) = f_{k-1}(\mathbf{x}) + \lambda h(\mathbf{x}, \bm{\alpha_k}) $ \\ 
}
\textbf{Return} $f(\mathbf{x}) = f_K(\mathbf{x})$ \\
\caption{Gradient boosting algorithm \autocite{FriedmanJeromeH.2001GfaA, B_hlmann_2007} }
\label{alg:gradient-boosting}
\end{algorithm}

\begin{algorithm}[htb!]
Initialise  $f_0(\mathbf{x}) = \arg \min_{\bm{\alpha_0}} \sum_{i=1}^N l(y_i, \mathbf{x}, \bm{\alpha_0})$ \\
\For{k = 1 : K}{
    For $i = 1,2,\dots N$, compute the gradient residual using $g_{ik} = - \left [ \frac{\partial l(y_i, f_{k-1}(\mathbf{x_i})) }{\partial f_{k-1}(\mathbf{x_i}) } \right ]$ \\
    Fit a decision tree to the targets $g_{ik}$ giving terminal leaves $R_{kj}$, $j = 1,2, \dots J_k$, where $J_k$ is the number of terminal leaves.\\ 
    For $j = 1,2, \dots J_k$, compute $\alpha_{jk} = \arg \min_{\bm{\alpha}} \sum_{\mathbf{x_i} \in R_{kj}} l(y_i, f_{k-1}(\mathbf{x_i}) + \bm{\alpha}) $ \\ 
    Update boosting trees with learning rate $\lambda$ $f_k(\mathbf{x}) = f_{k-1}(\mathbf{x}) + \lambda \sum_{j=1}^{J_k} \alpha_{kj} \mathbb{I} (\mathbf{x} \in R_{kj})  $ \\ 
}
\textbf{Return} $f_K(x)$ \\
\caption{Gradient boosting tree algorithm implemented in LightGBM \autocite{LightGBM,FriedmanJeromeH.2001GfaA, B_hlmann_2007} }
\label{alg:gradient-boosting-tree}
\end{algorithm}

\end{document}